\documentclass[aps,prd,twocolumn,nofootinbib,showpacs,superscriptaddress,floatfix]{revtex4-1}  
\usepackage{graphicx}  
\usepackage{dcolumn}   
\usepackage{bm}        
\usepackage{amssymb}   
\usepackage{amsmath}
\usepackage{comment}
\usepackage[hidelinks]{hyperref}
\usepackage{graphicx}
\usepackage{subfig}
\usepackage{color}
\usepackage{slashed}
\usepackage{url}
\usepackage{hyperref}
\usepackage{lipsum}

\definecolor{darkturq}{RGB}{0, 206,209}

\newcommand{\Eq}[1]{Eq.(\ref{#1})}

\newcommand{\Fig}[1]{Fig. \ref{#1}}

\newcommand{\xs}{\mathbf{x}}
\newcommand{\ib}{\mathbf{i}}
\newcommand{\xt}{\mathbf{x}_{\bot}}
\newcommand{\rsph}{r_{\text{sph}}}
\newcommand{\tsph}{t_{\text{sph}}}

\begin{document}
 
\author{Mark Mace}
\email{mark.mace@stonybrook.edu}
\affiliation{Physics and Astronomy Department, Stony Brook University, Stony Brook, NY 11973, USA}
\affiliation{Physics Department, Brookhaven National Laboratory, Bldg. 510A, Upton, NY 11974, USA}
\author{Niklas Mueller}
\email{n.mueller@thphys.uni-heidelberg.de}
\affiliation{Institut f\"{u}r Theoretische Physik, Universit\"{a}t Heidelberg, Philosophenweg 16, 69120 Heidelberg, Germany}
\author{S\"{o}ren Schlichting}
\email{sslng@uw.edu}
\affiliation{Department of Physics, University of Washington, Seattle, WA 98195-1560, USA}
\author{Sayantan Sharma}
\email{sayantans@quark.phy.bnl.gov}
\affiliation{Physics Department, Brookhaven National Laboratory, Bldg. 510A, Upton, NY 11973, USA}

\title{Non-equilibrium study of the Chiral Magnetic Effect from real-time simulations with dynamical fermions}

\begin{abstract}
We present a real-time lattice approach to study the non-equilibrium dynamics of vector and axial charges in $SU(N) \times U(1)$ gauge theories. Based on a classical description of the non-Abelian and Abelian gauge fields, we include dynamical fermions and develop operator definitions for (improved) Wilson and overlap fermions that allow us to study real-time manifestations of the axial anomaly from first principles. We present a first application of this approach to anomalous transport phenomena such as the Chiral Magnetic Effect (CME) and Chiral Separation Effect (CSE) by studying the dynamics of fermions during and after a $SU(N)$ sphaleron transition in the presence of a $U(1)$ magnetic field. We investigate the fermion mass and magnetic field dependence of the suggested signatures of the CME and CSE and point out some important aspects which need to be accounted for in the macroscopic description of anomalous transport phenomena.
\end{abstract}
\date{\today}

\maketitle
\section{Introduction}
Quantum anomalies are ubiquitous in nature and have lead to many fascinating phenomena in quantum field theories. One of the most prominent examples occurs in the context of electroweak baryogenesis, 
where the combined effects of anomalous baryon plus lepton number violation and the $C$ and $CP$ violation may provide an explanation for the observed matter/anti-matter asymmetry under suitable out-of-equilibrium 
conditions~\cite{Sakharov:1967dj,Cohen:1993nk,Rubakov:1996vz}. Even though analogous processes exist in QCD, where the conservation of the axial charge $j^{0}_{a}$ is anomalously violated locally, observing such effects is a 
more subtle issue as QCD by itself does not violate the discrete $P$ and $CP$ symmetries globally.

In recent years, a major discovery is that the combination of QCD and QED effects, expected to occur in a Quark-Gluon Plasma (QGP), can lead to new macroscopic manifestations of real-time quantum anomalies~\cite{Fukushima:2008xe}, which could potentially be observed in high-energy heavy-ion collision experiments \cite{Kharzeev:2007jp}. While several phenomena are presently being discussed in this context (for review, see e.g. \cite{Kharzeev:2015znc}), the basic idea can be summarized as follows: topological transitions such as sphalerons~\cite{Klinkhamer:1984di,Arnold:1987mh}, which are expected to occur frequently in the QGP \cite{Moore:2010jd,Mace:2016svc}, can induce a net axial charge asymmetry $j^{0}_{a}$ of light quarks which can fluctuate on an event-by-event basis. Even though this axial charge asymmetry cannot be observed directly, in the presence of the strong electromagnetic $\vec{B}$ field created in off-central heavy-ion collisions it can be converted into an electric current $\vec{j} \propto j^{0}_{a} \vec{B}$~\cite{Vilenkin:1980fu}. This phenomenon is called the Chiral Magnetic Effect (CME) (for review, see e.g. \cite{Kharzeev:2013ffa}) and can 
lead to observable consequences in heavy-ion experiments~\cite{Kharzeev:2007jp,Kharzeev:2010gr}.

Experimental searches for the CME are ongoing at RHIC and the LHC, and intriguing hints suggestive of the CME have been seen across different experiments~\cite{Abelev:2009ac,Abelev:2012pa,Adamczyk:2015eqo}. 
Based on the notion that the CME should lead to a separation of electric charges across the direction of the magnetic field \cite{Kharzeev:2007jp}, the focus of experimental searches has been to measure the effects 
of electric charge separation at early times by analyzing charge dependent azimuthal correlations in the final state~\cite{Voloshin:2004vk}. However, it turns out that conventional explanations in terms of background effects also exist for the proposed observables and so far it has been a challenge to disentangle signal and background~\cite{Schlichting:2010na,Schlichting:2010qia,Pratt:2010zn,Bzdak:2010fd,Hatta:2015hca,Huang:2015fqj,Khachatryan:2016got}.
Experimentally, this question will be addressed in the near future through a proposed isobar run at RHIC~\cite{Skokov:2016yrj}. By studying the variation of the charge separation signal for two isobars, this experimental program is specifically designed to separate magnetic field independent backgrounds from the genuine CME signal~\cite{Voloshin:2010ut,Chatterjee:2014sea,Deng:2016knn}. Of course, along with the dedicated experimental efforts, there is a simultaneous need for an improved theoretical 
understanding of the expected magnitude and features of possible CME signals~\cite{Skokov:2016yrj}.

Over the past few years, a variety of different theoretical approaches have been developed to investigate the real-time dynamics of anomalous transport phenomena such as the 
CME across different physical systems. In particular, this includes macroscopic descriptions in terms of anomalous hydrodynamics \cite{Son:2009tf,Sadofyev:2010pr,Hirono:2014oda,Yin:2015fca} as well as microscopic 
descriptions based on chiral kinetic theory at weak coupling~\cite{Son:2012wh,Stephanov:2012ki,Gao:2012ix} and holographic methods at strong coupling~\cite{Yee:2009vw,Amado:2011zx, Lin:2013sga,Iatrakis:2014dka,Iatrakis:2015fma}. 
Despite all these developments, significant uncertainties remain with regard to a quantitative theoretical description of the CME in heavy-ion collisions~\cite{Kharzeev:2015znc,Skokov:2016yrj}. Since the lifetime of the magnetic field is 
presumably very short\footnote{Even though the lifetime of magnetic field induced by the spectators is extremely short, there is a possibility that a large magnetic field 
can be induced the QCD medium, which may survive on a somewhat longer time scale~\cite{Gursoy:2014aka}. However, the spacetime evolution of the induced magnetic field crucially depends e.g. on the chemical composition of the plasma at very early times and so far no firm conclusions have been reached concerning its actual importance~\cite{Skokov:2016yrj}.} \cite{Skokov:2009qp,McLerran:2013hla,Tuchin:2014iua}, a dominant source of uncertainty is an incomplete understanding of the dynamics of axial and vector charges during the early time pre-equilibrium stage~\cite{Skokov:2016yrj}. In order to address precisely these uncertainties, we recently advocated the use of a classical-statistical lattice approach which is specifically devised to explore the real time dynamics in far-from-equilibrium situations~\cite{Mueller:2016ven}.

Classical-statistical lattice techniques are a commonly used tool in the study of far-from-equilibrium many body systems. In the context of high-energy heavy-ion collisions, a classical-statistical treatment of the 
early time dynamics can be systematically derived within the Color Glass Condensate effective field theory~\cite{McLerran:1993ni,McLerran:1993ka,Gelis:2010nm} in the weak-coupling limit $(\alpha_s \ll 1)$: Since the phase space 
occupancies of gluons are non-perturbatively large $f \sim 1/\alpha_s$ at initial times, quantum effects are suppressed by an additional power of $\alpha_s$ and the early time dynamics can be accurately described in 
terms of an ensemble of classical Yang-Mills fields~\cite{Berges:2013fga}. 

Over the past few years classical-statistical lattice techniques have been employed to study various aspects of the early time, non-equilibrium dynamics, starting with the initial state particle production~\cite{Krasnitz:1998ns,Lappi:2003bi,Schenke:2012fw,Schenke:2015aqa} towards the onset of the 
thermalization process~\cite{Berges:2013eia,Berges:2013fga,Berges:2012ev,Gelis:2013rba,Kurkela:2012hp}. While so far most works have focused on the dynamics of the gluon fields, which dominate the dynamics in the high-occupancy regime ($f\gg1$), first attempts have also been made to include dynamical 
fermions into the description of the early time non-equilibrium dynamics~\cite{Gelis:2005pb,Hebenstreit:2013qxa,Tanji:2015ata,Gelis:2015kya,Gelfand:2016prm,Tanji:2016dka}.

Based on a classical-statistical lattice gauge theory description of the bosonic degrees of freedom, the real-time quantum dynamics of fermions can be studied from first principles within this approach by numerically solving the operator Dirac equation \cite{Aarts:1998td,Aarts:1999zn,Mueller:2016aao,Kasper:2014uaa,Fukushima:2015tza}. While the approach itself is not new as similar techniques have been employed previously e.g. in the context of strong field QED~\cite{Mueller:2016aao,Kasper:2014uaa,Buividovich:2015jfa}, 
cold electroweak baryogenesis~\cite{Saffin:2011kc,Saffin:2011kn} or cold quantum gases~\cite{Kasper:2015cca}, we have achieved several improvements which allowed us for the first time to study the $3+1$ D dynamics of 
anomalous transport phenomena in $SU(N_c) \times U(1)$ theories~\cite{Mueller:2016ven}. 

Our paper provides a detailed followup study to the previous letter~\cite{Mueller:2016ven}. A detailed exposition of the theoretical formalism is provided in Sec.~\ref{sec:classical_qcd}, including for the first time a real-time formulation of overlap fermions with exact chiral symmetry on the lattice\footnote{During the final stages of preparing this manuscript, we became aware of an exploratory study using real-time overlap fermions in a QED-like theory~\cite{Buividovich:2016ulp}.}. We then present several new physics results on the real-time dynamics of axial charge production in Sec.~\ref{sec:B_Zero} and anomalous transport processes in Sec.~\ref{sec:CME}. Even though our present numerical studies are performed in a minimal setup of a 
single $SU(2)$ sphaleron transition in a constant external $U(1)$ magnetic field, they provide novel insights into the real-time dynamics of anomalous transport effects and serve as a first important step in extending this 
approach towards a more realistic description of high-energy heavy-ion collisions. A compact summary of our findings and future perspectives is presented in Sec.~\ref{sec:conclusion}. Supplementary information is provided in the following Appendices.

\section{Classical-statistical lattice gauge theory with dynamical fermions}
\label{sec:classical_qcd}
We first describe our setup to perform classical-statistical real-time lattice gauge theory simulations with dynamical fermions coupled simultaneously to non-Abelian $SU(N_c)$ and Abelian $U(1)$ gauge fields. Even 
though we will only consider the $SU(2) \times U(1)$ case in our simulations, the discussion is kept general in anticipation of future applications to the $SU(3) \times U(1)$  case relevant to heavy-ion physics. 
Our simulations are performed in $3+1$ dimensional Minkowski spacetime ($g^{\mu\nu}=\text{diag}(1,-1,-1,-1)$), and we will denote the spacetime coordinate $x^{\mu}$ as $(t,x,y,z)$.  

We employ temporal axial $(A^{t}=0)$ gauge and work in the Hamiltonian formalism of lattice gauge theory, first formulated by Kogut and Susskind \cite{Kogut:1974ag}, where time $t$ remains a continuous coordinate 
while the spatial coordinates $\xs=(x,y,z)$ are discretized on a lattice of size $N_x \times N_y \times N_z$ with periodic boundary conditions and lattice spacing $a_s$ along each of the three dimensions. We choose a compact U(1) gauge group, such that both the non-Abelian and Abelian gauge fields are represented in terms of the usual lattice gauge link variables $U_{\xs,i} \in SU(N)\times U(1)$, where 
$\xs \in \{0,\dots,N_x-1\} \times \{0,\dots,N_y-1\} \times \{0,\dots,N_z-1\}$ denotes the spatial position and $i=x,y,z$ the spatial Lorentz index. 

Since the classical-statistical lattice formulation for gauge fields has been extensively discussed in the literature (see e.g. \cite{Berges:2012ev}), we will focus on the practical realization of the fermion dynamics, noting that the foundations of the formalism have been laid out in \cite{Aarts:1998td,Kasper:2014uaa}. Since there are various complications with respect to the realization of continuum symmetries of fermions on the lattice, we have implemented two different discretization schemes for fermions in this work. We will first discuss the real-time lattice formulation with Wilson fermions and subsequently describe the real-time lattice formulation with overlap fermions. 

\subsection{Wilson Fermions in real time}
\label{sec:Wison_Fermion_Real_Time}
Our starting point for the real-time lattice formulation with dynamical Wilson fermions is the lattice Hamiltonian operator, which takes the general form\footnote{We omit explicit factors of the lattice spacing. Hence all definition are given in dimensionless lattice units.}~\cite{Greiner:1985ce}
\begin{eqnarray}
\label{eqn:WilsonHamiltonian}
\hat{H}_W &=& \frac{1}{2} \sum_{\xs} [\hat{\psi}^{\dagger}_{\xs}, \gamma^{0} \Big(-i\slashed{D}^{s}_{W} + m \Big) \hat{\psi}_\xs].  \\ \nonumber
\end{eqnarray}
Here the fermion fields obey the usual anti-commutation relations
 \begin{eqnarray}
\{\hat\psi^\dagger_{\xs,a},\hat\psi_{\mathbf{y},b}\}&=&\delta_{\xs,\mathbf{y}}\delta_{a,b}~,
\end{eqnarray}
where $a,b$ collectively stand for spin and color indices and $-i\slashed{D}^{s}_{W}$ denotes the tree-level improved Wilson Dirac operator
\begin{eqnarray}
-i\slashed{D}^{s}_{W}\hat{\psi}_\xs&=&\frac{1}{2} \sum_{n,i} C_{n}
\Big[ \Big(-i\gamma^{i}- n r_{w} \Big) U_{\xs,+ni} \hat{\psi}_{\xs+n\ib} \\ \nonumber
&+&2 n r_{w} \hat{\psi}_{\xs} - 
\Big(-i \gamma^{i}+n r_{w} \Big)U_{\xs,-ni} 
\hat{\psi}_{\xs-n\ib} \Big]\;.  \nonumber 
\end{eqnarray}
By $r_{w}$ we denote the Wilson coefficient and we introduced the following short hand notation for the connecting gauge links
\begin{eqnarray}
U_{\xs,+ni}=  \prod_{k=0}^{n-1}U_{\xs+k\ib,i}\;, \qquad U_{\xs,-ni}=  \prod_{k=1}^{n}U^{\dagger}_{\xs-k\ib,i}.
\end{eqnarray}
Based on an appropriate choice of the coefficients up to $C_{n}$ it is possible to explicitly cancel lattice artifacts $\mathcal{O}(a^{2n-1})$ in the lattice Hamiltonian. By choosing only $C_{1}=1$ and all other coefficients 
to vanish, one recovers the usual (unimproved) Wilson Hamiltonian, which is only accurate to $O(a)$. With the first two terms $C_{1}=4/3$ and $C_{2}=-1/6$ we can achieve an $O(a^3)$ (tree level) improvement, 
and by including also the third term $C_{1}=3/2\;, C_{2}=-3/10\;, C_{3}=1/30$ we get an $O(a^5)$ {tree level) improvement.\footnote{Note that our improvement procedure parallels that of Ref.~\cite{Eguchi:1983xr}. Alternatively one could follow the procedure detailed in Ref.~\cite{Sheikholeslami:1985ij}, leading to the appearance of the familiar Clover term.}

\subsubsection{Operator decomposition and real-time evolution}
While the gauge links $U_{\xs,i}$ are treated as classical variables, it is important to keep track of the quantum 
mechanical operator nature of the fermion fields. Evolution equations for the fermion operators are derived from the lattice Hamiltonian, as
\begin{equation}
\label{eq:DiracEquationW}
i \gamma^{0} \partial_t \hat{\psi}_x=(-i\slashed{D}^{s}_{W}+m)\hat{\psi}_x \;,
\end{equation} 
which can be solved on the operator level by performing a mode function expansion~\cite{Aarts:1998td,Kasper:2014uaa}. Considering for definiteness an expansion in terms of the eigenstates of the Hamiltonian at initial 
time ($t=0$) the mode function decomposition takes the form
\begin{eqnarray}
\label{eq:ModeDecomposition}
\hat{\psi}_\xs (t)=\frac{1}{\sqrt{V}} \sum\limits_{\lambda}\left( \hat{b}_{\lambda}(0)\phi^u_{\lambda}(t,\xs) +
\hat{d}^{\dagger}_{\lambda}(0)\phi^v_{\lambda}(t,\xs)  \right)\;,
\end{eqnarray}
where $\lambda=1,\cdots,2 N_c N_x N_y N_z$ labels the energy eigenstates and  $\hat{b}(0)/\hat{d}^{\dagger}(0)$ correspond to the (anti) fermion (creation) annihilation operators acting on the initial state 
$(t=0)$~\cite{Aarts:1998td,Kasper:2014uaa}. By construction the time dependence of the fermion field operator $\hat{\psi}$ is then inherent to the wave-functions $\phi^{u/v}_{\lambda}(t,\xs)$, 
whereas the operator nature of $\hat{\psi}$ only appears through the operators $\hat{b}(0),\;\hat{d}^{\dagger}(0)$ acting in the initial state. 

Since for a classical gauge field configuration the Dirac equation (\ref{eq:DiracEquationW}) is linear in the fermion operator, it follows from the decomposition in \Eq{eq:ModeDecomposition} 
that the wave-functions $\phi^{u/v}_{\lambda}(t,\xs)$ satisfy the same equation. One can then immediately compute the time evolution of fermion field operator by solving the Dirac equation for each of 
the $4 N_c N_x N_y N_z$ wave functions. We obtain the numerical solutions using a leap-frog discretization scheme with time step $a_{t}=0.02a_s$. 

In practice performing the decomposition in \Eq{eq:DiracEquationW} amounts to the diagonalization of the matrix
\begin{eqnarray}
\label{eq:MatrixDiagonalization}
\gamma^{0} \Big(-i\slashed{D}^{s}_{W} + m \Big) \phi^{u/v}_{\lambda}(0,\xs)= \pm \epsilon_{\lambda}  \phi^{u/v}_{\lambda}(0,\xs)\;,
\end{eqnarray}
at initial time, where $\epsilon_{\lambda} \geq m$ denotes the energy of single particle states. In the simplest case, where the gauge fields vanish at initial time, the eigenfunctions $\phi^u_{\lambda}$ 
correspond to plane wave solutions and can be determined analytically as discussed in App.~\ref{app:eigenmodes}. However, if we introduce a non-vanishing magnetic field $B$ at initial time 
(see Sec.~\ref{sec:u1_links}), this is no longer the case and we instead determine the eigenfunctions $\phi^u_{\lambda}$ numerically using standard matrix diagonalization techniques.\footnote{Despite the fact that well known analytic solutions exist in the continuum in the case of a constant homogenous magnetic field, we are not aware of an equivalent analytic solution to \Eq{eq:MatrixDiagonalization} on the lattice.}

\subsubsection{Initial conditions and operator expectation values}
When computing any physical observable, one has to evaluate the operator expectation values with respect to the initial state density matrix. We will consider for simplicity an initial vacuum state, 
characterized by a vanishing single particle occupancy of fermions and anti-fermions $n_\lambda^{u/v}=0$  yielding the following operator expectation values 
\begin{eqnarray}
\langle [\hat b^{\dagger}_{\lambda},\hat b_{\lambda'}]\rangle &=&+2 (n_\lambda^{u}-1/2)\delta_{\lambda,\lambda'}\\
\langle [\hat d_{\lambda},\hat d_{\lambda'}^{\dagger}]\rangle&=&-2(n_\lambda^{v}-1/2) \delta_{\lambda,\lambda'} 
\end{eqnarray}
whereas all other combinations of commutators vanish identically. Specifically for this choice of the initial state, the expectation values of a local operator $\hat{O}(t,\xs)$ involving a commutator of two fermion fields can be expressed according to
\begin{eqnarray}
\hat{O}(t,\xs)= \sum_{\mathbf{y}} O^{ab}_{\mathbf{x}\mathbf{y}}~\frac{1}{2}[\hat{\psi}^{\dagger}_{\mathbf{x},a}(t),\hat{\psi}_{\mathbf{y},b}(t)]
\end{eqnarray}
The expectation value of this bilinear form can be expressed according to
\begin{eqnarray}
\langle \hat{O}(t,\xs) \rangle &=& \frac{1}{V} \sum_{\lambda,\mathbf{y}} \Big[ \phi^{u \dagger}_{\lambda,a}(t,\xs) O^{ab}_{\mathbf{xy}} \phi^{u}_{\lambda,b}(t,\mathbf{y}) (n_\lambda^{u}-1/2) \nonumber \\
&& \quad - \phi^{v \dagger}_{\lambda,a}(t,\xs) O^{ab}_{\mathbf{xy}} \phi^{v}_{\lambda,b}(t,\mathbf{y})  (n_\lambda^{v}-1/2) \Big]\;.
\end{eqnarray}
as a weighted sum over the matrix elements of all wave-functions.

\subsubsection{Vector and axial currents}
We will consider vector $j^{\mu}_{v}$ and axial currents $j^{\mu}_{a}$ as our basic observables in this study. Since time remains continuous in the Hamiltonian formalism, vector and axial densities are defined 
in analogy to the continuum as
\begin{equation}
j^{0}_{v}(x)=\frac{1}{2}\langle [\hat \psi^\dagger_{x}, \hat \psi_{x}]\rangle\;, \qquad j^{0}_{a}(x)=\frac{1}{2}\langle [\hat \psi^\dagger_{x},  \gamma_{5} \hat \psi_{x} ] \rangle~.
\end{equation}
and no extra terms occur for the time-like components. However, this is different for the spatial components of the currents, where additional terms arise in the lattice definition.  By performing the variation of 
the Hamiltonian with respect to the Abelian gauge field, we obtain the spatial components of the vector currents according to 
\begin{eqnarray}
\lefteqn{j^{i}_{v}(x)=}\\
&\displaystyle\sum_{n,k=0}^{n-1}& \frac{C_{n}}{4} \Big \langle [\hat{\psi}^\dagger_{\xs-k\ib},\gamma^0 \Big(\gamma^{i}-in r_{w} \Big) U_{\xs-k\ib,ni}~\hat \psi_{\xs+(n-k)\ib}]\nonumber \\
&+& [\hat{\psi}^\dagger_{\xs+(n-k)\ib}, \gamma^0 \Big(\gamma^{i}+in r_{w} \Big)U_{\xs+(n-k)\ib,-ni}~\hat{\psi}_{\xs-k\ib}] \Big  \rangle \nonumber~.
\end{eqnarray}
Since the currents are derived from the improved Hamiltonian, these are by construction improved which is important for reducing discretization effects as we will discuss in more detail in the upcoming section.

Defining the axial currents requires a more careful analysis to recover the correct anomaly relations in the continuum limit. In order to fully appreciate this point, let us first recall that for a naive discretization of the fermion action (obtained e.g. by setting $r_w=0$) an unphysical cancellation of the anomaly takes place, which can be understood as a consequence of the doubling of fermion modes~\cite{Karsten:1980wd}. Hence the correct realization of the axial anomaly for Wilson fermions relies on lifting the degeneracy between doublers by introducing the Wilson term $(r_w\neq0)$, and achieving an effective decoupling of the fermion doublers in the continuum limit~\cite{Karsten:1980wd}. Defining the spatial components of the axial current as
\begin{eqnarray}
j^{i}_{a}(x)&=& \sum_{n,k=0}^{n-1} \frac{C_{n}}{4} \Big \langle [\hat{\psi}^\dagger_{\xs-k\ib},\gamma^0 \gamma^i \gamma_5~ U_{\xs-k\ib,ni}~\hat \psi_{\xs+(n-k)\ib}]  \nonumber \\
&+& [\hat{\psi}^\dagger_{\xs+(n-k)\ib}, \gamma^0 \gamma^i \gamma_5~ U_{\xs+(n-k)\ib,-ni}~\hat{\psi}_{\xs-k\ib}] \Big  \rangle.
\end{eqnarray}
it can easily be shown that the axial current for lattice Wilson fermions satisfies the exact relation
\begin{eqnarray}
\partial_{\mu} j^{\mu}_{a}(x) = 2 m \eta_{a}(x) + r_w  W(x),
\label{eqn:lattice_anomaly}
\end{eqnarray}
where $\partial_{i} j^{i}_{a}(x)=j^{i}_{a}(x)-j^{i}_{a}(x-i)$ and $\eta_{a}(x)$ denotes the pseudoscalar density
\begin{eqnarray}
\eta_{a}(x)=\frac{1}{2}\langle [\hat{\psi}^{\dagger}_x, i \gamma^{0} \gamma_{5}  \hat{\psi}_x] \rangle 
\end{eqnarray}
and $W(x)$ is the explicit contribution from the Wilson term
\begin{align}
W(x)= \sum_{n,i} \frac{n\cdot C_n }{4}\Big\langle [&\hat\psi^\dagger_{\xs}, i\gamma_5\gamma_0 \big( U_{\xs,+ni}\hat\psi_{\xs+n\mathbf{i}} -2\hat\psi_\xs \nonumber \\
&+ U_{\xs-n\mathbf{i},+ni}^\dagger \hat\psi_{\xs-n\mathbf{i}} \big) ] +\text{h.c.}\Big \rangle
\end{align}
Even though the lattice anomaly relation in \Eq{eqn:lattice_anomaly} may appear unfamiliar at first sight, it has been shown in the context of Euclidean lattice gauge theory, that the usual form is recovered in the 
continuum limit, where the Wilson term gives rise to a non-trivial contribution
\begin{eqnarray}
r_{w} W(x) \to -\frac{g^2}{8\pi^2} \text{Tr} F_{\mu\nu}(x) \tilde{F}^{\mu\nu}(x),
\end{eqnarray}
$F_{\mu\nu}$ being the field strength tensor and $\tilde{F}^{\mu\nu}=\frac{1}{2} \epsilon^{\mu\nu\rho\sigma} F_{\rho \sigma}$ is it's dual.
It can also be shown that the first deviations from the continuum limit appear as an odd function of $r_w$ and improved convergence can be achieved by averaging of positive and negative values of the Wilson 
parameter~\cite{Aoki:1983qi,De:2011xp}. Even though the generalization of these proofs to the non-equilibrium case is non-trivial, explicit numerical verification 
has been reported in \cite{Tanji:2016dka} and we will confirm this behavior in Sec.~\ref{sec:B_Zero} based on our own simulations.

\subsection{Overlap fermions in real time}
\label{sec:Overlap_Fermion_Real_Time}
\subsubsection{Constructing the Overlap Hamiltonian}
Wilson fermions break the chiral and anomalous $U_A(1)$ symmetries explicitly on the lattice. Explicit chiral symmetry is recovered only in the continuum limit for massless Wilson fermions\footnote{Note that mass renormalization effects can render this issue problematic, as a carful tuning of the Wilson bare mass is required in taking the correct continuum limit. However, since we will only consider the dynamics of fermions in a classical background field, such problems are absent in the simulations present in this work.}. With the improvement procedures for the Wilson fermions one can reduce the lattice artifacts responsible for chiral symmetry breaking, however it is still desirable to compare our results with a lattice fermion discretization where the chiral and continuum limits are clearly disentangled. Overlap fermions~\cite{Narayanan:1993ss,Neuberger:1997fp} have exact chiral and flavor symmetries on the lattice and the anomalous $U_A(1)$ symmetry can be realized even for a finite lattice spacing, analogous to the way it happens in the continuum. Even though we will demonstrate that within our simple setup one can obtain comparable results with improved Wilson and Overlap fermions, we point out that the real-time overlap formulation may be important for future real-time simulations that either go beyond classical background fields or involve truly chiral fermions.

We will now employ overlap fermions for real-time simulations of the anomaly induced transport phenomena. As we did for the Wilson fermions, we consider a Hamiltonian formulation which for massless overlap quarks,
\begin{eqnarray}
\label{eqn:OverlapOperatorh}
\hat{H}_{ov}=\frac{1}{2}\sum_\xs [\hat{\psi}^\dagger_\xs, \gamma_0 \big( -i \slashed{D}_{ov}^s \big) \hat{\psi}_\xs ]
\end{eqnarray}
Here $-i\slashed{D}_{ov}^s$ is the 3D spatial overlap Dirac operator given by
\begin{eqnarray}
\label{eqn:OverlapOperatorh}
-i\slashed{D}_{ov}^s=M\left(\mathbf{1}+ \frac{\gamma_0 H_W(M)}{\sqrt{H_W(M)^2}}\right)
\end{eqnarray}
and $H_W(M)$ is the original Wilson Hamiltonian kernel, defined in \Eq{eqn:WilsonHamiltonian} but with $C_n=0$ for $n\geq2$, and with the fermion 
mass $m$ being replaced by the negative of the domain wall height $M$, namely, 
\begin{eqnarray}
H_W(M)=\gamma_0 (-i\slashed{D}_W^s-M).
\end{eqnarray}
The domain wall height takes values $M \in (0,2]$. In Appendix~\ref{app:Ovhamiltonian} we derive the Hamiltonian for the first time in the overlap formalism. We note that it is assuring that this construction is in 
exact agreement with the ansatz for the overlap Hamiltonian for vector-like gauge theories first discussed in \cite{Creutz:2001wp}. Furthermore simulating massive overlap quarks is straightforward within this setup, 
which can be implemented by simply replacing 
\begin{eqnarray}
-i\slashed{D}^{s}_{ov} \rightarrow -i\slashed{D}^{s}_{ov}\left(1-\frac{m}{2M}\right)+m,
\end{eqnarray}
where $m$ is the quark mass we want to simulate.

The overlap Dirac matrix for massless quarks in three spatial dimensions, $\slashed{D}_{ov}^s$ satisfies the Ginsparg-Wilson relation~\cite{Ginsparg:1981bj},
\begin{equation}
\label{eqn:gwrelation}
\{\slashed{D}_{ov}^s,\gamma_5\}= -i \slashed{D}_{ov}^s\gamma_5\slashed{D}_{ov}^s~.
\end{equation}
Additionally the overlap Dirac operator is $\gamma_0$-hermitian, and satisfies a variant of \Eq{eqn:gwrelation},
\begin{equation}
\label{eqn:gwrelationg0}
\{\slashed{D}_{ov}^s,\gamma_0\}= -i \slashed{D}_{ov}^s\gamma_0\slashed{D}_{ov}^s~.
\end{equation}
As a consequence, it was shown in \cite{Creutz:2001wp} that the Hamiltonian commutes with the operator
\begin{equation}
\hat{Q}_5=\frac{1}{2}\displaystyle\sum_\xs \left[\psi^\dagger_\xs,\gamma_5\left(1-\frac{-i\slashed{D}_{ov}^s}{2}\right)\psi_\xs \right].
\end{equation}
This allows one to define $\hat{Q}_5$ as the axial charge within the Hamiltonian formalism, whose time evolution is given by the equation,
\begin{equation}
\label{eqn:q5evolution}
\frac{d \hat{Q}_5}{d t}=i[\hat{H}_{ov},\hat{Q}_5]+\frac{\partial \hat{Q}_5}{\partial t}~.
\end{equation} 
Since the first term in the right hand side of \Eq{eqn:q5evolution} is identically zero by construction, the time dependence of the axial charge density operator arises from the explicit real-time evolution of the matter fields in the definition of $\hat{Q}_5$. Hence in the real-time overlap formulation, the axial charge is generated exactly in the same way as in the continuum. While in \cite{Creutz:2001wp} the definition of the axial charge operator, $\hat{Q}_5$, is motivated from the symmetries of the overlap Hamiltonian, we show below how this definition arises naturally from the spatial integral of the time component of the axial current.

\subsubsection{Vector and axial currents in the overlap formalism }
Since the overlap operator has exact chiral symmetry on the lattice one can define chiral projectors which project onto fermion states with definite handedness. The left and the right-handed fermion fields can be defined in terms of lattice projection operators $ \check{P}_\pm$ as
\begin{eqnarray}
\psi_{R/L} =\frac{1}{2}(\mathbf{1} \pm \check{\gamma}_5)\psi \equiv \check{P}_\pm\psi,
\label{eqn:OverlapPsiProjector}
\end{eqnarray}
where $\check{\gamma}_5\equiv\gamma_5(\mathbf{1}+i\slashed{D}^s_{ov})$. In order to satisfy the Ginsparg-Wilson relation, the chiral projectors for the conjugate fields are then
\begin{eqnarray}
\psi^\dagger_{R/L}=\psi^\dagger \frac{1}{2}(\mathbf{1} \pm \gamma_5)\equiv \psi^\dagger P_\pm.
\label{eqn:OverlapPsiBarProjector}
\end{eqnarray}
Instead of following the approach to define currents from the variation of the Hamiltonian, we can define vector currents in analogy to the continuum by constructing these quantities in terms of the 
physical left and right-handed fermion modes~\cite{Gavai:2011np,Narayanan:2011ff}. Based on this approach, the vector current for overlap fermions in real-time is constructed as
\begin{eqnarray}
j_v^\mu&=&\frac{1}{2}\langle [ \hat{\psi}^\dagger_R,\gamma_0 \gamma^\mu \hat{\psi}_R ] \rangle+\frac{1}{2}\langle [ \hat{\psi}^\dagger_L,\gamma_0\gamma^\mu \hat\psi_L ] \rangle\nonumber \\
&=&\frac{1}{2}\langle [ \hat\psi^\dagger,\gamma_0\gamma^\mu \Big(\mathbf{1} - \frac{-i \slashed{D}_{ov}^s}{2} \Big) \hat\psi ] \rangle;
\end{eqnarray} 
similarly the axial current is
\begin{eqnarray}
j_a^\mu&=&\frac{1}{2}\langle [ \hat{\psi}^\dagger_R,\gamma_0 \gamma^\mu \hat{\psi}_R ] \rangle - \frac{1}{2}\langle [ \hat{\psi}^\dagger_L,\gamma_0\gamma^\mu \hat\psi_L ] \rangle \nonumber \\
&=&\frac{1}{2}\langle [ \hat\psi^\dagger, \gamma_0 \gamma^\mu \gamma_5 \Big(\mathbf{1}-\frac{-i \slashed{D}_{ov}^s}{2} \Big) \hat\psi ] \rangle ~.
\label{eqn:overlap_axial}
\end{eqnarray} 

\subsubsection{Numerical Implementation of the overlap operator}
The overlap Hamiltonian consists of a matrix sign function of $H_W(M)$, defined in \Eq{eqn:OverlapOperatorh}. The inverse square root of $H_W(M)^2$ can be expressed as a Zolotarev rational 
function~\cite{Edwards:1998yw,vandenEshof:2002ms,Chiu:2002eh,Gavai:2002uy},
\begin{equation}
\label{eqn:Zolotarevp}
\frac{1}{\sqrt{ H_W(M)^2}}=\sum_{l=1}^{N_\mathcal{O}}\frac{b_l}{d_l+ H_W(M)^2 }.
\end{equation}
To compute \Eq{eqn:Zolotarevp}, first we compute the coefficients $b_l$ and $d_l$ from the smallest and largest eigenvalues of $H_W(M)^2$~\cite{Chiu:2002eh}. Once the Zolotarev expansion coefficients $d_l$ are determined, we implement a multi-shift conjugate gradient solver 
to calculate the inverse of $d_l+H_W(M)^2$. The lowest and the highest eigenvalues for $H_W(M)^2$ are calculated using the Kalkreuter-Simma Ritz algorithm~\cite{Kalkreuter:1995mm} with $20$ restarts and a convergence 
criterion of $10^{-20}$. We find that taking $N_\mathcal{O}=20$ terms in the Zolotarev polynomial results in $| \text{sign}(H_ W )^ 2 - \mathbf{1}| < 10^{-9}$. We note that the lowest and highest eigenvalues of 
$H_W(M)^2$ are sensitive to the choice of the domain wall height $M$. We have chosen $M$ such that we obtain the best approximation to the sign function as well as the Ginsparg-Wilson relation. For the sphaleron 
configuration we studied in this work the optimal choice was $M\in[1.4,1.6)$ (see App.~\ref{app:convergence} for more details).

For the multi-shift conjugate gradient, the convergence of the conjugate gradient is determined by the smallest $d_l$, and the convergence criterion is set to $ |H_W(M)^2| - 1 < 10^{-16}$. For the largest lattice volumes that we consider in this study and for the single $SU(2)$ sphaleron gauge configuration to be introduced in Sec.~\ref{sec:su2_links}, the conjugate gradient algorithm reaches the convergence criterion before the maximum number of steps, which we choose to be 2000. We have also checked that the resultant overlap Dirac operator satisfies the Ginsparg-Wilson relation, and found this is satisfied to a precision of $\mathcal O(10^{-9})$. We have also carefully studied the $M$-dependent cut-off effects for the vector and axial-vector currents which we would illustrate in the subsequent sections as well as in the Appendix~\ref{app:convergence}. We find that the cut-off effects in the current operators are fairly independent of the choice of $M$ for $M\in[1.4,1.6)$.

Additionally, we have also implemented the overlap Hamiltonian in the presence an additional static $U(1)$ magnetic field to be introduced in Sec.~\ref{sec:u1_links}. For this, we include the $U(1)$ fields in the 
Wilson Hamiltonian in \Eq{eqn:OverlapOperatorh}. We find that the sign function is implemented to a precision of $10^{-9}$ and the overlap Dirac operator in this case satisfies the Ginsparg-Wilson relation 
to a precision of $10^{-8}$.

\subsection{Non-Abelian and Abelian gauge links}
Within the classical-statistical approach, the dynamics of non-Abelian and Abelian fields is usually determined self-consistently by the solution to the classical Yang-Mills and Maxwell equations. In particular, the presence of the fermionic currents in the equations of motion for the gauge fields leads to a back-reaction of fermions, which is naturally included in the approach~\cite{Aarts:1998td,Aarts:1999zn}. Even though it will be desirable to investigate 
such effects in the long run, in the present study we will limit ourselves to a simpler set-up. Instead of a self-consistent determination of the non-Abelian and Abelian gauge fields, we will treat both of them as classical background fields whose dynamics is a priori prescribed. 

\subsubsection{ $SU(2)$ gauge links}
\label{sec:su2_links}
Concerning the $SU(2)$ gauge links, the dynamics is chosen to mimic that of a sphaleron transition by constructing a dynamical transition between topologically distinct classical vacua. Starting from the trivial vacuum solution 
$U^{\text SU(2)}_{\mathbf{x},i}=1$ at initial time $t=0$, we construct a smooth transition to a topologically non-trivial vacuum $U^{\text SU(2),G}_{\mathbf{x},i}$ at time $t \geq \tsph$ through a constant chromo-electric field, corresponding to the shortest path in configuration space,
\begin{eqnarray}
E_{\mathbf{x},i}^{a} =   \left\{ \begin{matrix} 
           \frac{i}{g a_s \tsph}\text{log}_{SU(2)}\big(U^{\text SU(2),G}_{\mathbf{x},i}\big) &,& 0<t<\tsph\\
            0 &,& t > \tsph &  \end{matrix} \right. \nonumber
            \\
\end{eqnarray}
during which the gauge links are constructed according to
\begin{eqnarray}
U^{\text SU(2)}_{\mathbf{x},i}(t)= 
        \left\{ \begin{matrix} 
           \rm{e}^{-iga_s t E_{\mathbf{x},i}^{a}\frac{\sigma^{a}}{2}}U^{\text SU(2)}_{\mathbf{x},i}(0) &,& 0<t<\tsph \\
            U^{\text SU(2),G}_{\xs,i} &,& t > \tsph &  \end{matrix} \right.
\end{eqnarray}
Since the different classical vacua are related to each other by a gauge transformation, we can easily construct a topologically non-trivial vacuum solution
\begin{equation}
U^{\text SU(2),G}_{\mathbf{x},i}=G_\mathbf{x} G_{\mathbf{x}+\mathbf{i}}^{\dagger}.
\label{eqn:vacuumconstruction}
\end{equation}
by specifying a gauge transformation $G_x$ with a non-zero winding number. Based on the usual parametrization of the $SU(2)$ gauge group,
\begin{equation}
G_\mathbf{x}=\alpha_0(\mathbf{x})\mathbf{1}+i\alpha_a(\mathbf{x})\sigma^a~,
\end{equation}
the coordinates $\alpha_{a}(\mathbf{x})$ of the gauge transformation on the group manifold are obtained by a distorted stereographic projection of the lattice coordinates $\mathbf{x}=(x,y,z)$, which has a non-zero Brouwer degree. By virtue of our construction detailed in App.~\ref{app:mapping}, the sphaleron transition profile (i.e all points that map away from the trivial point $G_\xs=1$) is localized on a scale $\rsph$, which we will refer to as the characteristic size scale of the sphaleron. 

\subsubsection{$U(1)$ gauge links}
\label{sec:u1_links}
With regard to the Abelian gauge links we have chosen to implement a homogenous magnetic field $\vec B= B \hat z$ along the z-direction. Since on a periodic lattice the magnetic flux $q a_s^2 B N_x N_y$ is quantized in 
units of $2\pi$~\cite{AlHashimi:2008hr}, a spatially homogenous magnetic field cannot be varied continuously and we have chosen to keep the magnetic field constant as function of time. By choosing the $U(1)$ components 
of the gauge links according to \cite{Bali:2011qj},
\begin{eqnarray}
\label{eqn:magbc}
U^{\text U(1)}_{\xs,x} &=& \left\{ \begin{matrix} 
	e^{i a_s^2 q B N_x y} &,& x=N_x-1 \\ 
	1 &,& \text{otherwise} &  \end{matrix} \right. \\
U^{\text U(1)}_{\xs,y} &=& 	e^{-i a_s^2 q B n_x} 
\end{eqnarray}
with $U_{\xs,z}=1$ and $a_s^2 q B= \frac{2\pi n_B}{N_x N_y}$ we can then realize different magnetic field strength by varying the magnetic flux quantum number $n_B$. 

\section{Sphaleron transition \& real-time dynamics of axial charge production in $SU(N)$ }\label{sec:B_Zero}
We now turn to the results of our simulations and first study the dynamics of axial charges during a sphaleron transition in the absence of electro-magnetic fields ($B=0$). Since the realization of the axial anomaly on the 
lattice is non-trivial a first important cross-check is to verify that the continuum version of the anomaly relation
\begin{eqnarray}
\partial_{\mu} j^{\mu}_a(x)= 2m \eta_{a}(x) -2\partial_{\mu}K^{\mu}(x)\;,
\label{eqn:anomaly}
\end{eqnarray}
where $\partial_{\mu}K^{\mu}(x)=\frac{g^2}{16\pi^2} \text{tr} F_{\mu\nu}\tilde F^{\mu\nu}$ denotes the divergence of the Chern-Simons current, is correctly reproduced in our simulations. If we focus on the volume integrated 
quantities 
\begin{eqnarray}
J_a^0(t)=\int d^3 \xs \; j_a^0(t,\xs)
\end{eqnarray}
the net axial charge $J_a^0$ can be directly related to the Chern-Simons number difference, according to
\begin{eqnarray}
\Delta J_a^0(t)=-2 \Delta N_{CS}(t)\,,
\label{eqn:integrated_anomaly}
\end{eqnarray}
which changes by an integer amount over the course of the sphaleron transition. Specifically, for the topological transition constructed in Sec.~\ref{sec:su2_links}, $\Delta N_{CS}(t\geq \tsph)=-1$ and one expects 
$\Delta J_a^0(t)=2$ units of axial charge to be created during the transition.

\begin{figure}
  \includegraphics[width=.95\linewidth]{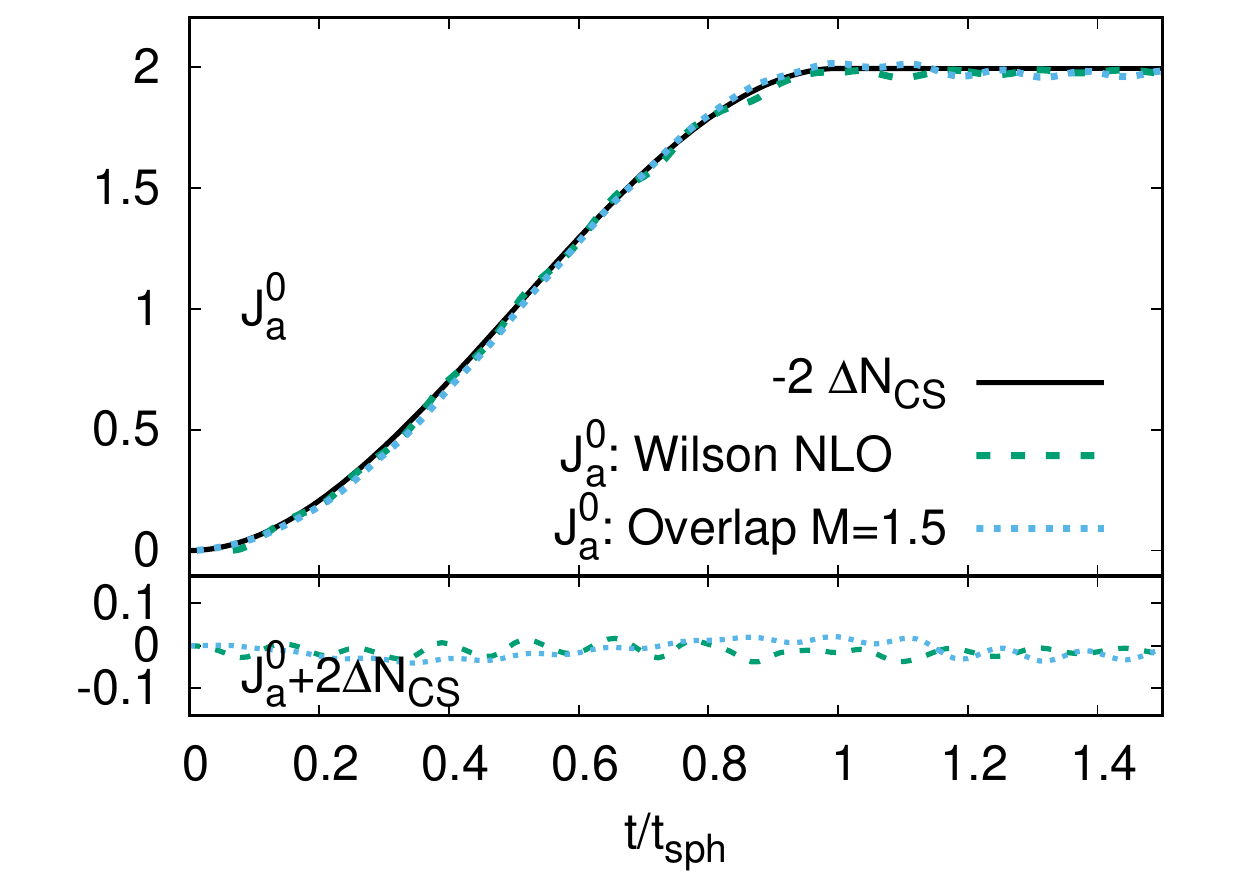}
\caption{A comparison of the net axial charge generated during a sphaleron transition for improved Wilson (NLO) fermions with $m \rsph=1.9 \cdot 10^{-2}$ versus massless overlap fermions on a $16^3$ lattice. Top: The net axial charge for both discretizations accurately tracks $\Delta N_{CS}$ due to the sphaleron transition. Bottom: Deviations from \Eq{eqn:integrated_anomaly} are shown.}
  \label{fig:HMS_COMP}
  \end{figure}

Simulation results for the real-time evolution of the net axial charge $J_a^0(t)$ are compactly summarized in Fig.\ref{fig:HMS_COMP}, where we compare results obtained for massless overlap fermions on a $16^3$ spatial 
lattice with the results obtained for light Wilson fermions ($m\rsph=1.9\cdot10^{-2}$). Since the typical size scale of the sphaleron $\rsph$ and duration of the sphaleron transition $\tsph$ are the only dimensionful 
parameters in this case, in the following all spatial and temporal coordinates will be normalized in units of $\rsph$ and $\tsph$ respectively; if not stated otherwise we employ $\tsph/\rsph=3/2$.

Since we employ a fermionic vacuum as our initial condition, the axial charge is zero initially, as there are no fermions present. As the sphaleron transition takes place fermions are dynamically produced and an axial imbalance 
is created. By comparing the evolution of $J_a^0(t)$ with that of the Chern-Simons number, extracted independently from the evolution of the gauge fields\footnote{We use an an $O(a^2)$ improved lattice definition described 
in detail in \cite{Moore:1996wn,Mace:2016svc}.}, it can be clearly seen that the global version of the anomaly relation in \Eq{eqn:integrated_anomaly}, is satisfied to good accuracy.

Concerning the comparison of different fermion discretizations, we find that the results for improved Wilson fermions (next to leading order) agree nicely with the ones obtained in the overlap formulation. However, we strongly emphasize 
that the operator improvements for Wilson fermions are essential to achieve this level of agreement on the relatively small $16^3$ lattices. If in contrast one was to consider unimproved Wilson fermions, much finer lattices 
are needed to correctly reproduce the continuum anomaly and we refer to App.~\ref{app:convergence} for further performance and convergence studies.

Even though our present results are obtained for a single smooth gauge field configuration, an important lesson can be inferred for upcoming studies on more realistic gauge fields. Since the computational cost of the 
simulations scales as $\propto N_x^2 N_y^2 N_z^2$, simulations on fine lattices are often prohibitively expensive and it is therefore of utmost importance to employ improved fermionic operators in real-time lattice simulations with dynamical fermions.

\begin{figure}
  \includegraphics[width=.95\linewidth]{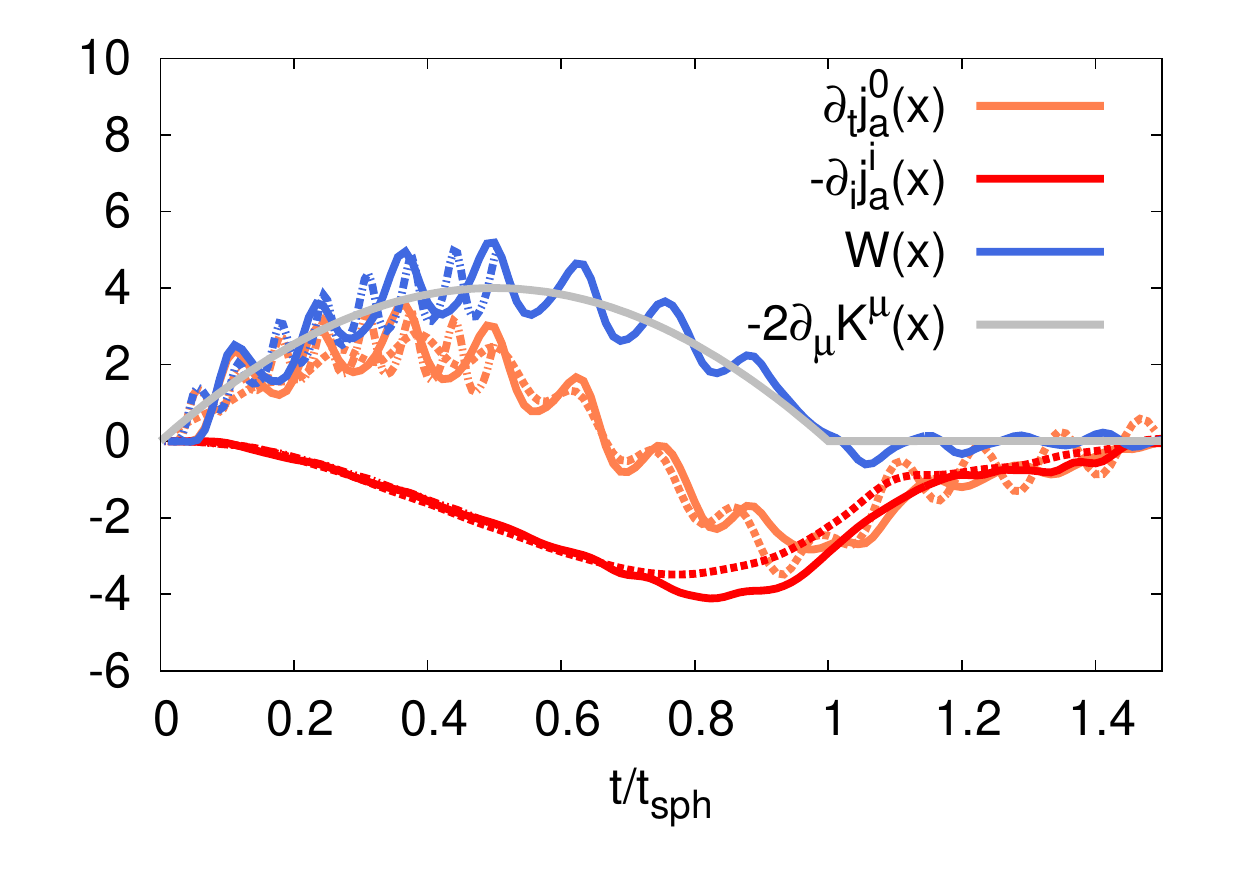}
  \caption{The local anomaly budget at the center of the sphaleron transition using improved Wilson (NLO) and overlap fermions. The solid, dash-dotted, and dotted lines represent data for improved Wilson (NLO) on a $16^3$ 
  lattice, $32^3$ lattice, and overlap fermions on a $16^3$ lattice respectively. The gray line represent the local derivative of the Chern Simons current, $-2 \partial_\mu K^\mu$. }
  \label{fig:localbudget}
  \end{figure}

Based on the excellent agreement obtained between different lattice and continuum results for volume integrated quantities, we can now proceed to study the microscopic dynamics of axial charge production in more detail. 
In Fig.~\ref{fig:localbudget} we present a breakup of the different contributions, $\partial_t j^{0}_{a},\partial_{i}j^{i}_{a}$ and $-2\partial_{\mu}K^{\mu}$, to the local anomaly budget (c.f. \Eq{eqn:anomaly}) evaluated at the center $(x,y,z)=(N_x/2,N_y/2,N_z/2)$ of the sphaleron transition profile. We have kept the volume fixed in units of $\rsph$ and to compare quantities between different lattice spacings and different fermion discretizations we have scaled the observables by appropriate powers of $\rsph$. Besides the rate of increase of the axial charge density $\partial_t j^0_{a}$, a significant fraction of the anomaly budget is 
compensated by the divergence of the axial current $\partial_{i}j^{i}_{a}$, signaling the outflow of axial charge from the center to the edges of the transition region. Hence, even though an axial charge imbalance is 
dominantly produced in the center of a sphaleron, axial charge redistributes as a function of time and the axial imbalance at the center again decreases towards later times. 

As discussed in Sec.~\ref{sec:Wison_Fermion_Real_Time}, the lattice anomaly relation for Wilson fermions is realized through the non-trivial continuum limit of the Wilson term $W(x)$ also depicted in Fig.~\ref{fig:localbudget}. 
Indeed, the evolution of the Wilson term $W(x)$ follows that of the evolution of divergence of the Chern-Simons current $-2\partial_{\mu}K^{\mu}$, albeit superseded by fast oscillations. However, the oscillations average 
out in both space and time yielding a faster convergence for time and/or volume averaged quantities. It also re-assuring that the comparison of the results for almost massless Wilson and chiral overlap fermions shows good 
overall agreement, although minor deviations remain on the presently available lattice sizes.

\begin{figure}
  \includegraphics[width=.95\linewidth]{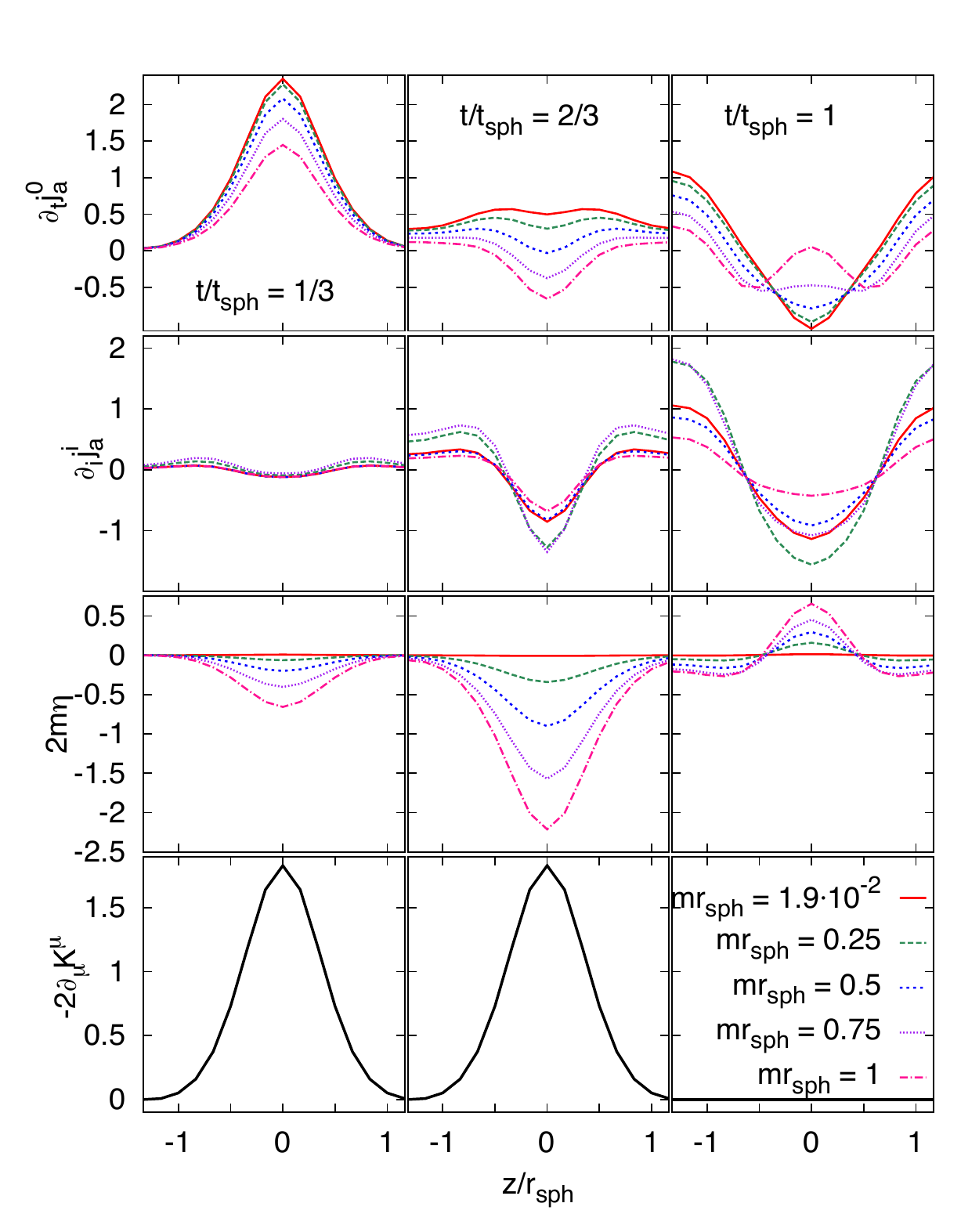}
  \caption{One dimensional profiles of the contributions to the anomaly equation for different masses in units of $r_\text{sph}^{-1}$. As can be seen, the rate of axial charge density production at the center of the sphaleron 
  is reduced due to axial currents carrying charge away and, in the case of a finite quark mass, by the pseudoscalar density, signaling chirality changing fermion-fermion interactions.}
  \label{fig:profiles_B0}
\end{figure}

\subsection{Quark mass dependence} 
\label{sec:nomag_quarkmass}
So far we have analyzed the non-equilibrium dynamics of axial charge production for (almost) chiral fermions. We will now vary the quark mass to investigate the effects of explicit chiral symmetry breaking on axial charge 
production. Before we turn to our physical results a technical remark is in order. Since we find that for Wilson fermions cut-off effects are more pronounced for larger values of the quark mass, we performed $r_w$ averaging 
of our results, i.e. we performed real-time evolutions with Wilson parameters $r_w=\pm 1$ respectively and calculated observables by averaging the results over each value of $r_w$. Based on this procedure, a compact summary 
of our results for massive fermions is compiled in Fig.~\ref{fig:profiles_B0}, showing freeze-frame profiles of the local anomaly budget for different values of the quark mass. Different panels show profiles of the (four) 
divergence of the Chern-Simons current $-2\partial_{\mu}K^{\mu}$, the pseudoscalar density $\eta$, the divergence of the axial current $\partial_{i}j^{i}_{a}$ and the time derivative of the local axial charge density 
$\partial_{t} j^{0}_{a}$,  along one of the spatial directions according to
\begin{align}
\partial_{\mu}K^{\mu}(z,t)=\frac{g^2}{8\pi^2} \int d^2\xt~E_{i}^{a}(x)B_{i}^{a}(x)\;,
\end{align}
and similarly for the other components at three different times $t/\tsph=1/3,\;2/3,\;1$ of the sphaleron transition. Different curves in each panel correspond to the results obtained for different values of the fermion mass  
ranging from almost massless quarks $m\rsph=1.9 \cdot 10^{-2}$ to intermediate values of $m\rsph=1$. 

Starting with the dynamics at early times ($t/\tsph=1/3$), the time derivative of the axial charge density shows a clear peak at the center corresponding to the creation of a local imbalance due to the sphaleron transition. 
While for almost massless quarks $m\rsph=1.9 \cdot 10^{-2}$, the rate of axial charge production $\partial_{t} j^{0}_{a}$ is approximately equal to the divergence of the Chern-Simons current $-2\partial_{\mu}K^{\mu}$, for heavier quark masses a significant fraction of the local anomaly budget is balanced by the contribution of the pseudoscalar density $2 m \eta$ resulting in a smaller rate of axial charge production, both locally and globally. 

Once a local imbalance of axial charge is created at the center, axial currents $j^{i}_a$ with a negative (positive) divergence $\partial_{i}j^{i}_a$ at the center (edges) develop and contribute an outflow of the axial charge 
density away from the center. Even though the divergence of the Chern-Simons current $-2\partial_{\mu}K^{\mu}$ remains positive at times $t/\tsph=2/3$, its contribution to the axial charge  production rate $j^{0}_{a}$ at 
the center is largely compensated by the outward flow of axial currents $\partial_{i}j^{i}$. In particular, for massive quarks ($m\rsph>1/2$), the combined effects of axial charge dissipation due to a large pseudoscalar 
density $2 m \eta$ and outflowing currents $\partial_{i}j^{i}$ lead to a depletion of axial charge at the center ($\partial_{t} j^{0}_{a}<0$) even though the sphaleron transition is still in progress. 

Subsequently at even later times, axial charge continues to spread across the entire volume leading to a depletion of axial charge at the center and an increase towards the edges. In the case of massive quarks, the 
pseudoscalar density contributes towards the dissipation of axial charges, and the global imbalance $J^{0}_{a}$ decreases significantly as a function of time. Our simulations clearly point to the importance of including 
such dissipative effects due to a finite quark masses, and we will further elaborate on their influence on the dynamics of axial and vector charges in Sec.~\ref{sec:CME_magnetic_mass}. 

\section{Chiral magnetic effect \& Chiral magnetic wave in $SU(N) \times U(1)$}
\label{sec:CME}
\begin{figure*}
\label{sfig:ma}
  \includegraphics[width=.9\textwidth]{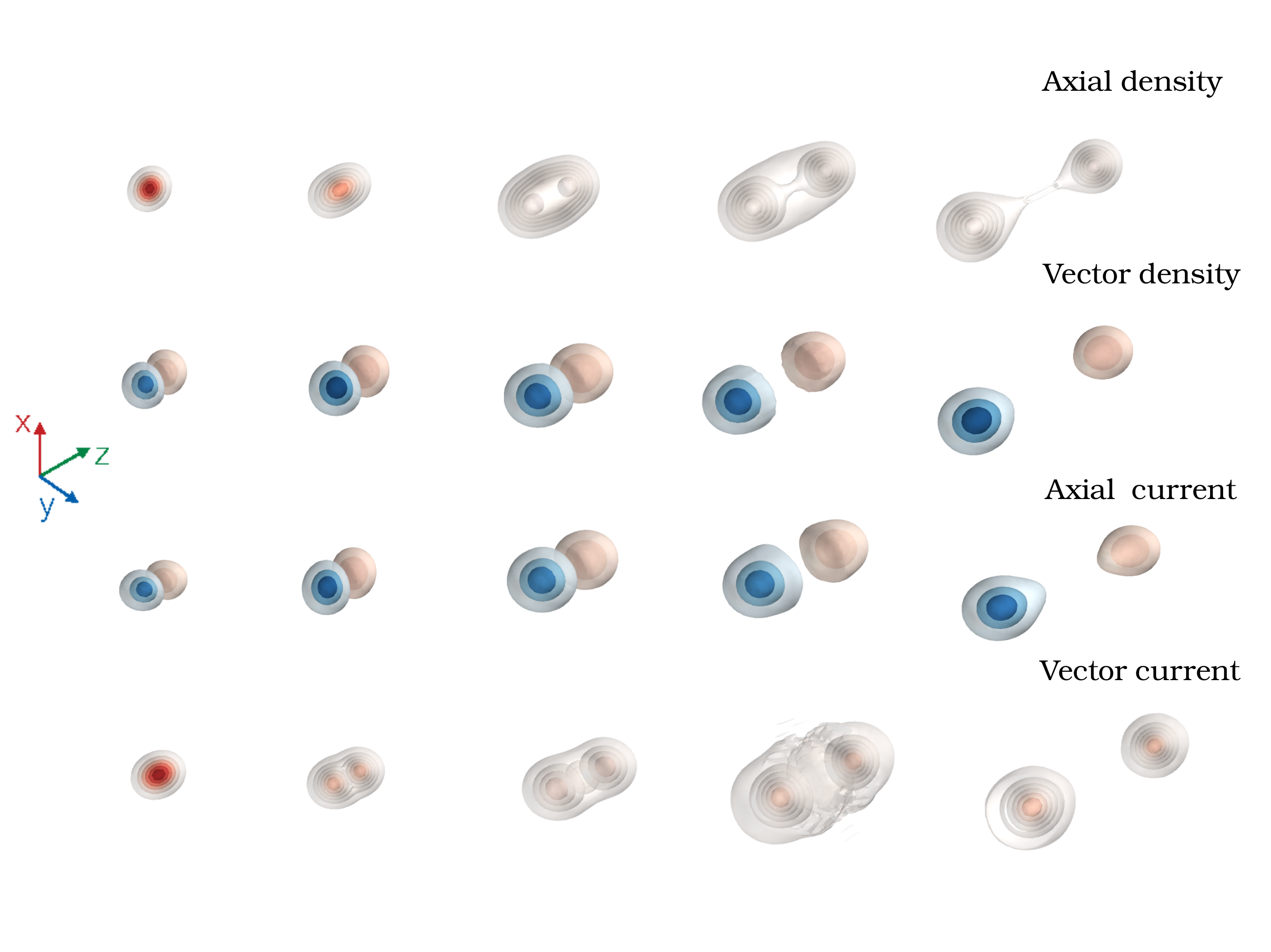}%
\caption{\label{Profiles} Profiles of the axial and vector densities and currents at different times of the real-time evolution for fermions with $m \rsph=1.9 \cdot 10^{-2}$ for strong magnetic fields $q B \rsph^{-2}=7.0$ at times $t/\tsph=0.6,0.9,1.1,1.3,1.6$.}
\label{fig:profilesB9}
\end{figure*}

We now turn to investigate the real-time dynamics of fermions during a sphaleron transition in the presence of a strong (Abelian) magnetic field.  Simulations are performed on larger $24\times 24 \times 64$ lattices 
with improved Wilson fermions. We consider a homogenous magnetic field $B$ in the $z$ direction (see Sec.~\ref{sec:u1_links}) and prepare the initial conditions as a fermionic vacuum in the presence of the magnetic field. Since the Abelian magnetic field introduces a non-trivial coupling between the dynamics of vector and axial charges due to the Chiral Magnetic Effect (CME) and Chiral Separation Effect (CSE)
~\cite{Kharzeev:2010gd}, the $SU(N)\times U(1)$ system exhibits interesting dynamics which we addressed previously in \cite{Mueller:2016ven}. Below we significantly expand upon our earlier results, concerning in particular 
the quark mass and magnetic field dependence of the dynamics.  Before we address these points in more detail, we will briefly illustrate the general features of the dynamics of vector and axial charges based on simulations 
for light quarks $m\rsph=1.9\cdot10^{-2}$ in a strong magnetic field $q B \rsph^{-2}=7.0$.

The basic features of the dynamics of vector and axial charges are compactly summarized in Fig.~\ref{fig:profilesB9}, showing three dimensional profiles of the axial and vector charge ($j^{0}_{a/v}$) and current ($j^{z}_{a/v}$) 
densities at different times $(t/\tsph=0.6,0.9,1.1,1.3,1.6)$ during and after a sphaleron transition. As discussed in the previous section, the $SU(N)$ sphaleron transition leads to the creation of an axial imbalance observed 
at early times in the top panel of Fig.~\ref{fig:profilesB9}. However, in the presence of the $U(1)$ magnetic field, the generation of an axial charge imbalance is now accompanied by the creation of a vector current along the 
magnetic field direction (CME), which can be observed in the bottom panel of Fig.~\ref{fig:profilesB9}. Clearly the spatial profile of the vector current follows that of the axial charge distribution as expected from the 
constitutive relation $j^{z}_{v} \propto j^{0}_{a} B^z$ for the Chiral Magnetic Effect. 

As seen in the second panel of Fig.~\ref{fig:profilesB9} the vector current leads to a separation of vector charges along the direction of the magnetic field at early times. Over the timescale of the sphaleron transition, 
positive (red) and negative (blue) charges accumulate at the opposites edges of the sphaleron transition region and give rise to a dipole-like structure of the vector charge distribution. Due to the Chiral Separation Effect (CSE), the presence of a local vector charge imbalance at the edges in turn induces an axial current which is depicted in the third panel of Fig.~\ref{fig:profilesB9} and leads to a separation of axial charge along 
the direction of the magnetic field. Ultimately the interplay of CME and CSE lead to formation of a Chiral Magnetic Wave, associated with the coupled transport of vector and axial charges along the direction of the 
magnetic field which can be observed at later times in Fig.~\ref{fig:profilesB9}. 

Specifically for light fermions in the presence of a strong magnetic field, the emerging wave packets of axial charge and vector current are strongly localized and closely reflect the spacetime profile of the sphaleron. However, as we will see 
shortly this is no longer necessarily the case for heavier fermions or weaker magnetic fields. We also note that in our present setup, the dynamics at late times is somewhat trivial as the outgoing shock-waves are 
effectively propagating into the vacuum. While in a more realistic scenario the number of sphaleron transitions at early times is presumably still of $\mathcal{O}(1)$~\cite{Mace:2016svc}, the chiral shock-waves are 
created from and move through a hot plasma and it will be interesting to observe how the subsequents dynamics is altered by further interactions with the constituents of the plasma.

Before we analyze the anomalous transport dynamics in more detail, we briefly comment on the comparison of Wilson and Overlap discretizations in the $SU(2)\times U(1)$ case. In order to perform a quantitative comparison 
of our results with different fermion discretizations, we will focus on the longitudinal profiles of vector and axial charge densities defined as
\begin{eqnarray}
j^{0}_{a/v}(t,z)=\int d^2\xt ~ j^{0}_{a/v}(t,\xt,z).
\label{eqn:long_profile}
\end{eqnarray}
Our results for somewhat smaller magnetic field strength $qB=3.5 \rsph^{-2}$ are compared in Fig.~\ref{fig:profiles_discret}, showing freeze-frame profiles of the longitudinal vector and axial charge distribution at three 
different times $t/\tsph=0.34,1,1.67$.  We observe a striking level of agreement between Wilson and Overlap results. Only at late times minor deviations between different discretizations become visible. However, at this point 
finite volume effects also start to become significant on the smaller $16\times 16 \times 32$ lattices employed for this comparison.

\begin{figure}
\includegraphics[width=.98\linewidth]{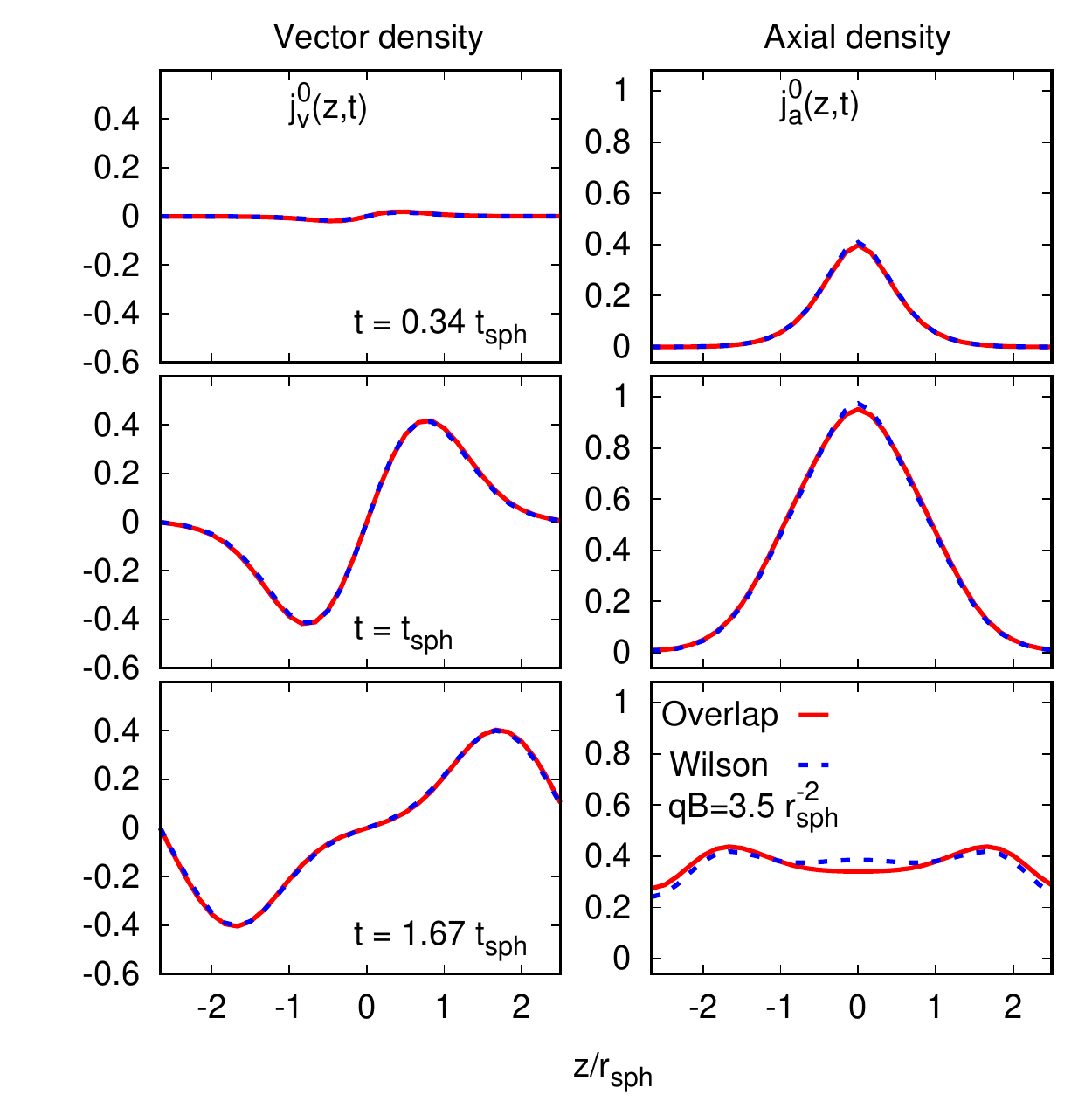}%
\caption{Comparison of longitudinal profiles of the vector (left) and axial (right) charge densities for improved Wilson (NLO) fermions and overlap fermions with masses $m \rsph=1.9 \cdot 10^{-2}$ in an external magnetic field $qB=3.5\rsph^{-2}$ at times $t/t_\text{sph}=0.34,\;1,\;1.67$ (top to bottom).
}
\label{fig:profiles_discret}%
\end{figure}

\subsection{Magnetic field dependence \& comparison to anomalous hydrodynamics}%
\begin{figure}
\includegraphics[width=0.98\linewidth]{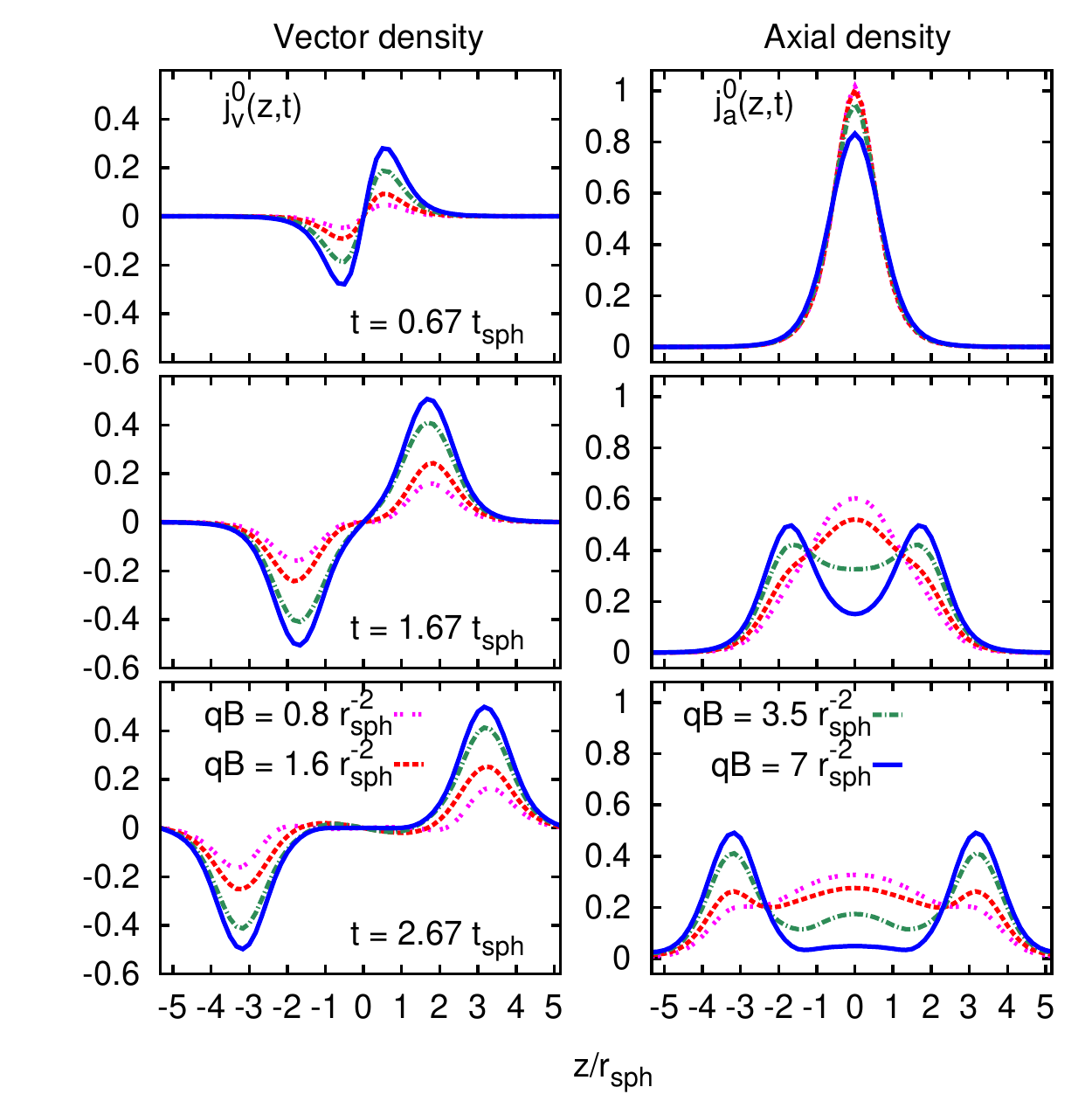}
\caption{Longitudinal profiles of the vector (left) and axial (right) charge density for different magnetic fields $qB$ in units of $\rsph^{-2}$ and for $m \rsph=1.9 \cdot 10^{-2}$ at times $t/t_\text{sph}=0.67,\;1.67,\;2.67$ (top to bottom).}
\label{fig:profiles_B}
\end{figure}

We will now investigate in more detail the magnetic field strength dependence of these anomalous transport phenomena. Even though the basic features of the dynamics of vector and axial charges observed in Fig.~\ref{fig:profilesB9} in the strong field limit remain the same for all values of the magnetic field considered in our study, some interesting changes occur when the magnitude of the magnetic field, $qB$, becomes comparable to the size of the inverse sphaleron radius squared, $\rsph^{-2}$, which is the other physical scale in our simulations.

Before we turn to the discussion of our simulation results, it is useful to first discuss how the magnetic field dependence enters in a macroscopic description in anomalous hydrodynamics~\cite{Son:2009tf}. In anomalous 
hydrodynamics the dynamics of vector and axial currents (in the chiral limit) is uniquely determined by the (anomalous) conservation of the (axial) vector currents
\begin{eqnarray}
\partial_{\mu}j^{\mu}_{v}=0\;, \qquad \partial_{\mu}j^{\mu}_{a}=-2\partial_{\mu}K^{\mu}\;, 
\end{eqnarray}
once the constitutive relations for the currents are enforced. In the ideal limit the constitutive relations take the form~\cite{Son:2009tf}
\begin{equation}
j^{\mu}_{v,a}=n_{v,a} u^{\mu} +\sigma^B_{v,a} B^{\mu}\;,
\label{eqn:constitutive_relation}
\end{equation}
and the magnetic field dependence enters only via the explicit $B$ dependence of the transport coefficient $\sigma^{B}_{v/a}$. In the weak field regime ($qB\ll \rsph^{-2}$) the conductivity is typically independent 
of the magnetic field and the CME/CSE currents are linearly proportional to the magnetic field $B$. In contrast in the strong field limit ($qB\gg \rsph^{-2}$), the conductivity of a free fermi gas becomes 
$\sigma^{B}_{v/a}=n_{a/v}/B$~\cite{Fukushima:2008xe} for a unit charge and the late time dynamics of vector and axial currents admits a simple analytic solution \cite{Mueller:2016ven}
\begin{align}
j^{0}_{v,a}&(t>t_{\rm{sph}},z) = \nonumber\\
&\quad \frac{1}{2} \int_{0}^{t_{\rm{sph}}} dt' ~\Big[ S\big(t',z-c (t-t')\big)  \mp S\big(t',z+c (t-t')\big) \Big]
\label{eqn:hydro_solution}
\end{align}
where $S(t,z)= -\frac{g^2}{8 \pi^2} \int d^2 x_{\bot} \text{Tr}~F^{\mu\nu} \tilde{F}_{\mu\nu}$ reflects the spacetime profile of the sphaleron transition. Most remarkably, the solution in \Eq{eqn:hydro_solution} shows explicitly that the anomalous transport dynamics becomes independent of the strength of the magnetic field $B$ in the strong field limit. However, this asymptotic scenario is unlikely to be realized in real-world 
experiments and it is hence important to understand the real-time dynamics of vector and axial charges beyond such simple asymptotic solutions.

Our simulation results for different magnetic field strength $q B \rsph^{2} =0.8,\;1.6,\;3.5,\;7.0$ are presented in Fig.~\ref{fig:profiles_B}, which shows the longitudinal profile of vector and axial charges densities $j^{0}_{a/v}(z,t)$ defined in \Eq{eqn:long_profile} 
for various times during and after the sphaleron transition. Even though the production of axial charge $j_{a}^{0}(z,t)$ during the transition ($t<\tsph$) is not altered significantly, the subsequent propagation of the chiral 
shock-waves is clearly affected by the strength of the magnetic field. While for the largest value of $q B\rsph^2=7.0$, the magnetic field can be interpreted as dominating over all other scales and the late time dynamics is 
accurately described by the asymptotic solution to anomalous hydrodynamics in \Eq{eqn:hydro_solution}, significant deviations from the asymptotic behavior occur for smaller values of $ q B\rsph^2=0.8,\;1.6,\;3.5$. Specifically, 
one observes from Fig.~\ref{fig:profiles_B} that a smaller CME current is induced for smaller values of the magnetic field, resulting in a reduced height of the vector charge peaks; in contrast the propagation velocities and profiles 
of the vector charge distribution are unaffected within this range of parameters.

Since a smaller amount of vector charge imbalance in turn leads to a reduction of the induced axial currents related to the CSE, clear differences emerge for the distribution of axial charges at later times. While for 
strong magnetic fields essentially all of the axial charge is subject to anomalous transport away from the center, a significant fraction of axial charge remains at the center for weaker magnetic field. Considering for 
instance the curves for $q B \rsph^{2} =1.6$, the axial charge distribution at later times can be thought of as a superposition of the free ($B=0$) distribution and the Chiral Magnetic Wave contributing clearly visible peaks at the edges.

\begin{figure}
\includegraphics[width=0.95\linewidth]{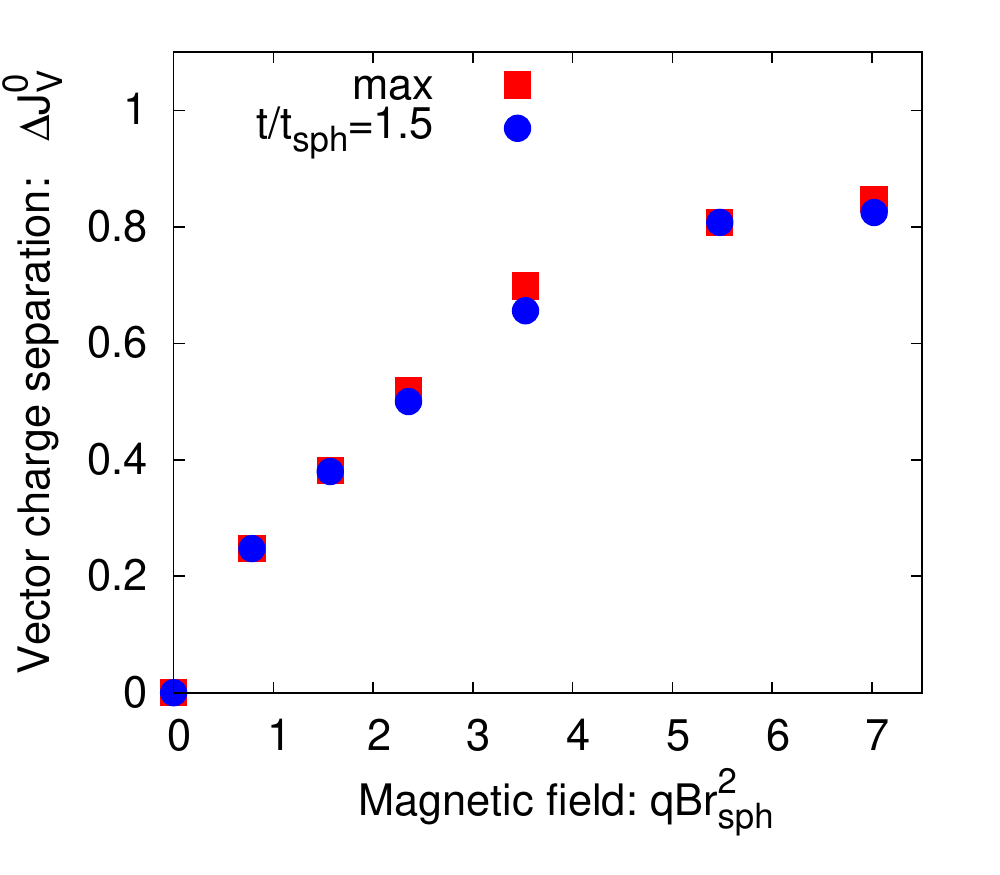}
\caption{Vector charge separation $\Delta J^{0}_{v}$ as a function of the magnetic field strength $qB$ in units of $\rsph^{-2}$.}
\label{fig:SepBDep}
\end{figure}

One can further quantify the magnetic field dependence by extracting the amount of vector charge separation achieved for different magnetic field strength. More precisely, we compute  
\begin{eqnarray}
\Delta J_{v}^{0}(t)=\int_{z\geq 0} dz ~j_{v}^{0}(t,z)\;,
\end{eqnarray}
corresponding to integrated the amount of vector charge contained in one of the oppositely charged wave-packets in Fig.~\ref{fig:profiles_B}. Simulation results for the magnetic field dependence of the charge separation 
signal are presented in \Fig{fig:SepBDep}, where different symbols correspond to the value of $\Delta J_{v}^{0}(t)$ at $t=3/2 \tsph$ and respectively the maximum value of $\Delta J_{v}^{0}(t)$ observed over the entire 
simulation time. In accordance with the expectation that the CME current is linearly proportional to the magnetic field strength in the weak field regime, one observes an approximately linear rise of the charge 
separation signal at smaller values of the magnetic field strength $qB \lesssim 4/\rsph^{2}$. In contrast for larger magnetic fields, the amount of vector charge separation begins to saturate, asymptotically 
approaching unity in the strong field limit.

\begin{figure}
 \includegraphics[width=.95\linewidth]{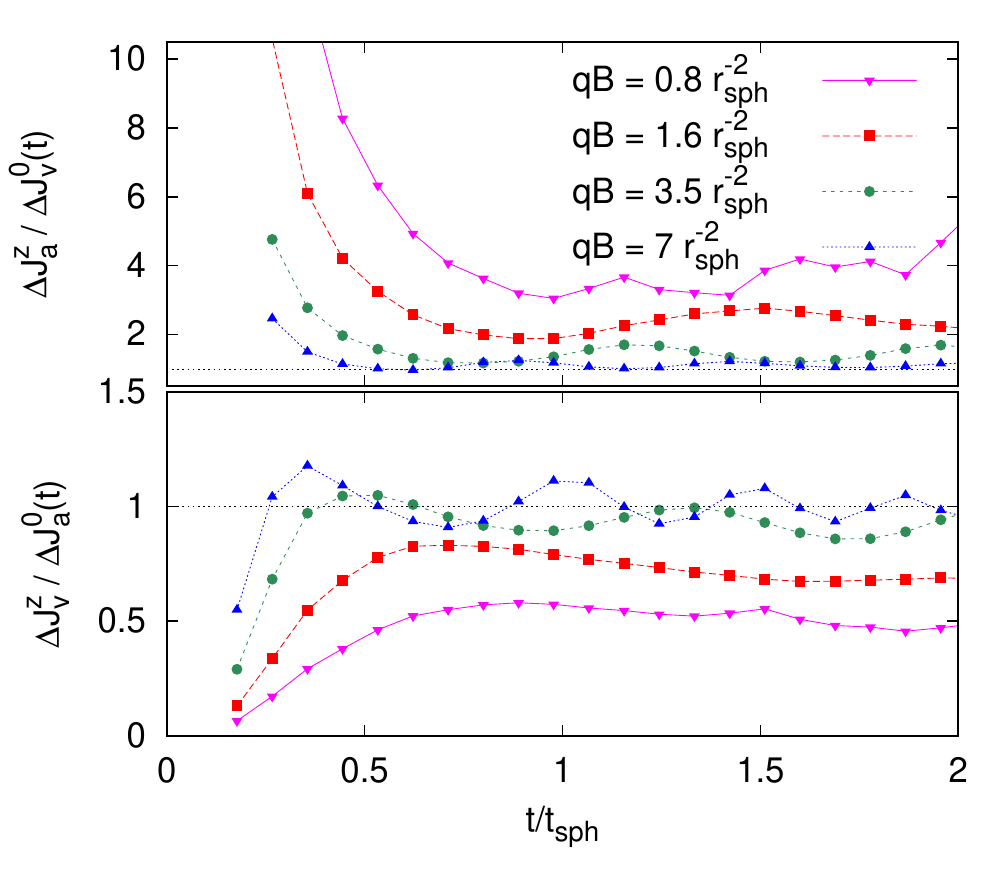}
  \caption{\label{fig:jvzDIVja0} (Top panel) Ratio between the axial current along the magnetic field and the electric charge (CSE) as a function of time for different magnetic field strength $qB$ in units of $\rsph^{-2}$. Bottom: Ratio between the electric current and the axial charge (CME).
  }
\end{figure}

Within our microscopic real-time description we can also attempt to verify directly to what extent the constitutive relations in \Eq{eqn:constitutive_relation} -- assumed in a macroscopic description in anomalous 
hydrodynamics -- are satisfied throughout the dynamical evolution of the system. In order to perform such a comparison, we extract the vector and axial charge $\Delta J_{a/v}^{0}(t)$ as well as the corresponding current 
densities $\Delta J_{a/v}^{z}(t)$ for the left- and right moving wave packets, and investigate the following ratios of net currents to net charges
\begin{eqnarray}
C_{{\rm CME}}(t)=\frac{\Delta J_{v}^{z}(t)}{\Delta J_{a}^{0}(t)}\;, \qquad C_{{\rm CSE}}(t)=\frac{\Delta J_{a}^{z}(t)}{\Delta J_{v}^{0}(t)}\;.
\end{eqnarray}
If one assumes the validity of the constitutive relations in \Eq{eqn:constitutive_relation}, one can immediately verify that both $C_{{\rm CME}}$ and $C_{{\rm CSE}}$ tend towards unity in the strong field limit~\cite{Fukushima:2008xe}. In contrast, the weak field regime constitutive relations take the form $\Delta J_{v/a}^{z} \propto  (\Delta J_{a/v}^{0})^{1/3} qB$ at low temperatures and $\Delta J_{v/a}^{z} \propto  (\Delta J_{a/v}^{0}) qB$ at high temperatures. Even though the ratios $C_{{\rm CME}}$ and $C_{{\rm CSE}}$ are no longer time independent constants in this limit, their numerical values are significantly smaller than unity and decrease as a function of axial/vector charge density \cite{Fukushima:2008xe}.

Our results for these ratios are presented in Fig.~\ref{fig:jvzDIVja0}, where we show the time evolution of $C^{{\rm eff}}_{{\rm CME}}$ and $C^{{\rm eff}}_{{\rm CSE}}$ for four different values of the magnetic field strength. Irrespective of the strength of the magnetic field one observes the same characteristic behavior of $C^{{\rm eff}}_{{\rm CME}}$ characterized by a rapid rise towards an approximately constant behavior at later times. In contrast for $C^{{\rm eff}}_{{\rm CSE}}$, the axial current $J_{a}^{z}$ also receives a contribution from the outflow of axial charge that is independent of the vector charge density $J^{0}_{v}$. Since the vector charge imbalance $J^{0}_{v}$ is initially small, this contribution dominates over the anomalous transport contribution at early times. Hence the current ratio $C^{{\rm eff}}_{{\rm CSE}}$ approaches its asymptotic value from above and can also exhibit asymptotic values larger than unity for small field strength.

Quantitatively the values observed for $C^{{\rm eff}}_{{\rm CME}}$ ($C^{{\rm eff}}_{{\rm CSE}}$) at later times are close to the strong field limit for $qB=3.5,7$ and slightly smaller (larger) for $qB=0.8,1.6$ and it is also important to point out that the initial build 
up of the CME and CSE currents occurs on a shorter time scale for larger magnetic field strength. Oscillations around the constant value are also clearly visible at late times and the oscillation frequency again depends 
strongly  on the strength of the magnetic field. However we can presently not exclude the possibility that the oscillations at late times are due to residual finite volume effects in our simulations and we will therefore 
not comment further on this behavior. 

While the results in Fig.~\ref{fig:jvzDIVja0} nicely confirm the approximate validity of constitutive relations at late times, it is also striking to observe that vector (CME) and axial (CSE) currents are not created 
instantaneously from the local imbalance of axial or vector charges. Conversely the results in Fig.~\ref{fig:jvzDIVja0} serve as a clear illustration of the retarded response and strongly suggest that, in order to describe 
the dynamics on shorter time scales, macroscopic descriptions should be modified to account for a finite relaxation time of anomalous currents. In the context of anomalous hydrodynamics, a natural way to include such 
effects is to follow the example of Israel and Stewart \cite{Israel:1979wp} by promoting the anomalous contribution to the currents to a dynamical variable $\xi^{\mu}_{v/a}$ that relaxes to the constitutive value 
$\sigma^B_{v/a} B^{\mu}$ on a characteristic time scale $\tau_{v/a}$. Since in high-energy heavy-ion collisions the lifetime of the magnetic field is presumably very short, it appears that the introduction of a 
finite relaxation time could indeed have quite dramatic effects. Hence it would also be important to understand more precisely which elementary processes determine the relevant time scale for the anomalous 
relaxation times. However, this question is beyond the scope of the present work.

\begin{figure}
\includegraphics[width=.98\linewidth]{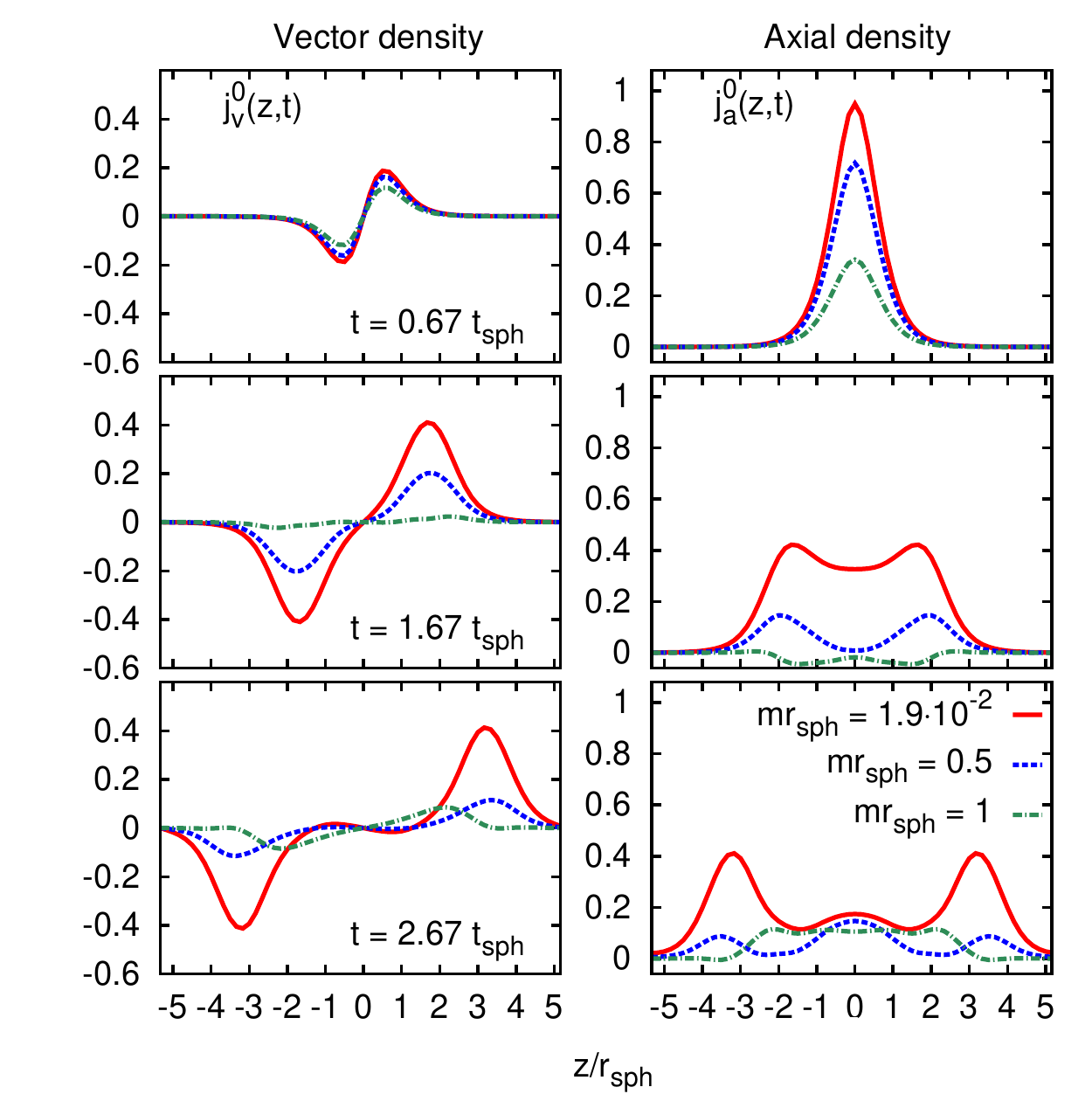}%
\caption{Longitudinal profiles of the vector (left) and axial (right) charge densities for different fermion masses in units of $\rsph^{-1}$ at times $t/t_\text{sph}=0.67,\;1.67,\;2.67$ (top to bottom).}
\label{fig:profilesmasses}%
\end{figure}

\subsection{Effects of finite Quark Masses}
\label{sec:CME_magnetic_mass}
We discussed in Sec.~\ref{sec:nomag_quarkmass} how explicit chiral symmetry breaking due to finite quark masses can significantly alter the production of an axial charge imbalance. We will now investigate in more detail the effects of 
explicit chiral symmetry breaking on the subsequent dynamics, characterized by the anomalous transport of axial and vector charges in the presence of a background magnetic field. Our results for different fermion 
masses are compactly summarized in Fig.~\ref{fig:profilesmasses}, where we show again the longitudinal profiles of vector and axial charge densities at different times during and after the sphaleron transition.  While 
the simulations are performed with improved Wilson fermions for a relatively large magnetic field strength, $qB\rsph^2=7.0$, we vary the masses from almost chiral fermions to fermions with large masses of the order of the inverse sphaleron size, 
$m\rsph=1$, where dissipative effects clearly become important on the time scales of interest. 

In accordance with the discussion in Sec.~\ref{sec:nomag_quarkmass} one observes from Fig.~\ref{fig:profilesmasses} that for heavier fermions $(m\rsph=0.5,1)$ the production of an axial charge imbalance at early times 
($t/\tsph=0.67$) is suppressed compared to the almost massless case $m\rsph=1.9 \cdot 10^{-2}$. Since the anomalous vector currents are locally proportional to the axial charge imbalance, a similar suppression 
of the vector charge density of heavier fermions $(m\rsph=0.5,1)$ can also be observed at early times ($t/\tsph=0.67$). Over the course of the evolution, drastic differences in the distribution of vector and axial 
charges emerge between light and heavy fermions. One clearly observes from Fig.~\ref{fig:profilesmasses}, how at times $t/\tsph=1.67,2.67$ the overall amount of axial and vector charge separation is strongly suppressed 
for larger values of the fermion mass $(m\rsph=0.5,1)$.  Moreover, as one would naturally expect for massive charge carriers, it is also evident from Fig.~\ref{fig:profilesmasses} that the propagation velocity of the chiral magnetic shock-waves decreases for larger values of the quark mass.

\begin{figure}
\includegraphics[width=.95\linewidth]{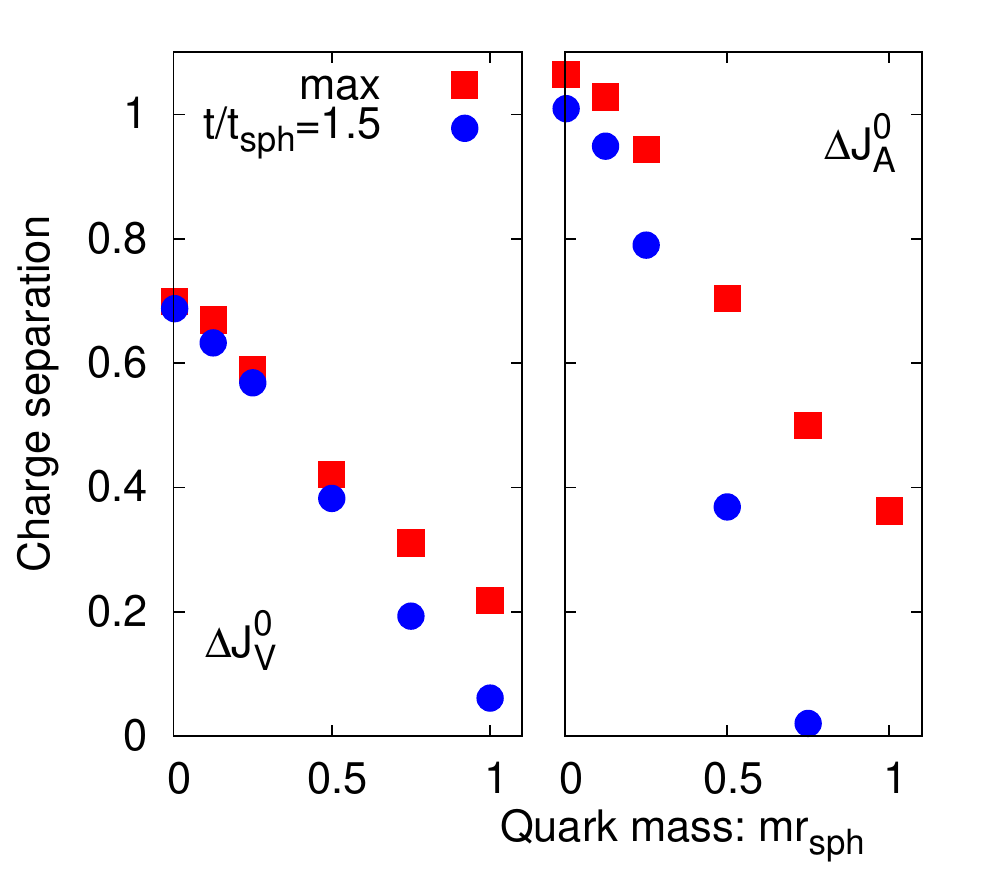}%
\caption{Vector (left) and axial (right) charge separation for different quark masses in units of $\rsph^{-1}$. The red points denote the maximum amount of charge separation during the entire real-time evolution; the blue points denote the amount of charge separation at a fixed time, $t/\tsph=1.5$, shortly after the sphaleron transition.
}
\label{fig:massdep}%
\end{figure}

In order to further quantify the quark mass dependence of the anomalous transport effects, we follow the same procedure outlined in Sec.~\ref{sec:nomag_quarkmass} and extract the vector and axial charge separation. Our results for the 
amount of vector/axial charge separation $\Delta J^{0}_{v/a}$ are presented in Fig.~\ref{fig:massdep} as a function of the quark mass. Different symbols in Fig.~\ref{fig:massdep} correspond to the vector/axial charge separation 
observed at a fixed time $t/\tsph=1.5$ and respectively the maximum value throughout the simulation ($0 \leq t/\tsph \leq 3$). Most strikingly, one observes from Fig.~\ref{fig:massdep} that clear deviations from the (almost) massless case emerge already for rather modest values of the quark mass. One finds that, for example for $m\rsph=0.25$, the observed vector charge separation signal is readily reduced by approximately $30 \%$. Considering 
even heavier quarks up to $m\rsph=1$, the vector charge separation signal almost disappears completely as dissipative effects dominate the dynamics. 

In view of the significant mass dependence observed in our simulations it would be interesting to compare our microscopic simulation results at finite quark mass to a macroscopic description of anomalous transport. However, 
we are presently not a aware of a macroscopic formulation that properly includes the effects of explicit chiral symmetry breaking. Even though mass effects might be small for phenomenological applications \cite{Hirono:2014oda,Yin:2015fca,Jiang:2016wve} in the light $(u,d)$ quark sector, they appear to be highly relevant with regard to the phenomenological description of the CME in the strange quark sector. Based on our results in Fig.~\ref{fig:massdep}, we expect a significant 
reduction of the possible CME signals for strange quarks, such that overall the situation may be closer to a two-flavor scenario~\cite{Kharzeev:2010gr}.

\section{Conclusions \& Outlook}
\label{sec:conclusion}
We presented a real-time lattice approach to study non-equilibrium dynamics of axial and vector charges in the presence of non-Abelian and Abelian fields. Even though the approach itself is by now well known 
and established in the literature, we pointed out several improvements related to the choice of the fermion discretization which are important to achieve a reliable description of the dynamics of axial charges 
in particular. Specifically, we pointed out that the use of tree-level improvements and $r$-averaging for the Wilson operator are essential to accelerate the convergence to the continuum limit and produce physical 
results on available lattice sizes. We also discussed the advantages and disadvantages of using overlap fermions in real-time lattice simulations and, to the best of our knowledge, performed the first real-time 3+1D 
lattice simulations with dynamical fermions with exact chiral symmetry. 

Based on our real-time non-equilibrium formulation, we studied the dynamics of axial charge production during an isolated sphaleron transition in $SU(2)$ Yang-Mills theory and explicitly verified that the axial anomaly recovered is satisfied to good accuracy at finite lattice spacing for both improved Wilson and overlap fermions. Beyond the dynamics for light fermions, we also investigated dissipative effects due to finite quark mass and reported how the emergence of a pseudoscalar density leads to a significant reduction of the axial charge imbalance created. Even though at present the sphaleron transition in the 
background gauge field configuration was constructed by hand and does not satisfy the equations of motion for the non-Abelian gauge fields, we emphasize that approximations of this kind made within our exploratory 
study can be relaxed in the future without any drawbacks on the applicability of our real-time lattice approach.

By introducing a constant magnetic field, we subsequently expand our simulations to a $SU(2)\times U(1)$ setup to study the real-time dynamics of anomalous transport processes such as the Chiral Magnetic and Chiral Separation Effect. We showed how the interplay of CME and CSE lead to the formation of a chiral magnetic shock-wave and demonstrated explicitly the dynamical separation of vector charges along the 
magnetic field direction. We also investigated in detail the quark mass and magnetic field dependence of these anomalous transport effects. Most importantly, we showed that the amount of vector charge separation 
created during this process is linearly proportional to the magnetic field strength (at small $qB$) and decreases rapidly as a function of the quark mass. Even though for light $(u,d)$ flavors, such quark mass effects 
are most likely negligible over the typical time scales of a heavy-ion collision, the situation is different with regard to strange quarks, where it appears necessary to take these effects into account in a 
phenomenological description. Since in contrast to the vector current the axial current is not conserved, it would be extremely important to investigate how creation and dissipation of axial charges, which are accurately described within our microscopic framework, can be accounted for within a macroscopic description. On a similar note, we also studied the onset of the CME and CSE currents and reported first evidence for a finite relaxation time of vector and axial currents. Even though a finite 
relaxation time may have important phenomenological consequences, given the short lifetime of the magnetic field in high-energy heavy-ion collisions, it is presently unclear which microscopic processes determine the 
relevant time scale and we intend to return to this issue in a future publication.
Our simulations were performed for an isolated sphaleron transition (see Sec. II C 1), allowing us to clearly observe non-perturbative generation and transport of axial charges in a topologically non- trivial background. However, the results presented in this paper can only serve as a qualitative benchmark of the real-time dynamics of anomalous transport effects. In a more realistic scenario one expects the quantitative behavior of anomalous transport to be modified through further interactions with the constituents of the plasma, and it will be in- teresting to explore these effects in more detail in the future by performing analogous studies on more realistic gauge field ensembles. 

Despite the fact that our present simulations of anomalous transport phenomena were performed in a drastically simplified setup, our work provides an important step towards a more quantitative theoretical understanding 
of the CME and associated phenomena in high-energy heavy-ion collisions. Since the life time of the magnetic field in heavy-ion collisions is short, it is important to understand the dynamics of anomalous transport 
during the early time non-equilibrium phase.  However, as we pointed out, the theoretical techniques developed in this work can be used to the address open questions in this context within a fully microscopic description 
of the early time dynamics. In the future it will be important to extend these studies to include more realistic gauge configurations and a spacetime dependent magnetic field in order to address important phenomenological issues. Besides the applications to high-energy nuclear physics, the theoretical approach advocated in this paper has a large variety applications e.g. in the study of cold electroweak baryogenesis~
\cite{Saffin:2011kc,Saffin:2011kn}, strong field QED~\cite{Mueller:2016aao}, or cold atomic gases~\cite{Kasper:2015cca}. In this context, the technical developments achieved in this work should also be valuable and we are 
looking forward to explore further applications of our ideas.

\section{Acknowledgements}
We thank J\"{u}rgen Berges, Dmitri Kharzeev, Jinfeng Liao, Larry McLerran, Raju Venugopalan, and Ho-Ung Yee for useful discussions and comments. We are supported in part by the U.S. Department of Energy under Grant No. DE-SC0012704 (M.M., Sa.S.), DE-FG88-ER40388 (M.M.), DE-FG02-97ER41014 (So.S.), and by the Studienstiftung des Deutschen Volkes and by the Deutsche Forschungsgemeinschaft Collaborative Research Centre SFB 1225 (ISOQUANT) (N.M.), and by the U.S. Department of Energy, Office of Science, Office of Nuclear Physics, within the framework of the Beam Energy Scan Theory (BEST) Topical Collaboration (M.M.). Sa.S. thanks the Institute for Nuclear Theory at the University of Washington for its hospitality and the Department of Energy for partial support during the completion of this work. This research used resources of the National Energy Research Scientific Computing Center, a U.S. Department of Energy Office of Science User Facility supported under Contract No. DE-AC02-05CH11231. Part of this work was performed on the computational resource ForHLR Phase I funded by the Ministerium f\"{u}r Wissenschaft, Forschung und Kunst Baden-W\"{u}rttemberg and the Deutsche Forschungsgemeinschaft. Additional numerical calculations were also performed using the USQCD clusters at Fermilab.

\appendix

\section{Eigenmodes of the Dirac Hamiltonian in the helicity basis}
\label{app:eigenmodes}
In this appendix we derive the eigenmodes for non-interacting fermions in the helicity basis by diagonalizing the Dirac Hamiltonian for Wilson and overlap fermions. 

We begin by taking the gamma matrices in the Dirac representation. In the absence of gauge fields ($U=\mathbf{1}$) the eigenfunctions of the Wilson and overlap Dirac equation can be written in the plane wave basis. The spatial momenta and effective mass term for the improved Wilson fermions in this basis are
\begin{eqnarray}
p^w_i&=&\sum_n\frac{C_n}{a_s}\text{sin}(n a_s q_i) \nonumber \\
m^w_{\text{eff}}&=&m+ \sum_{n,i} \frac{2 n C_n}{a_s} r_w \text{sin}^2(\frac{n a q_i}{2})
\end{eqnarray}
and similarly for massless overlap fermions\footnote{For overlap, we always take $r_w=1$ and the Wilson improvement coefficients $C_1=1$,$C_n=0$ for $n > 1$}
\begin{eqnarray}
p^{ov}_i&=&M \frac{p^w_i}{s} \nonumber \\
m^{ov}_{\text{eff}}&=&M\left(1+\frac{p_5}{s}\right)
\end{eqnarray}
where 
\begin{eqnarray}
q_i&=&\frac{2 \pi n_i}{N_i},~~n_i\in 1,...,N_i-1 \nonumber \\
p_5&=&-M+ \sum_{i} \frac{2}{a_s} \text{sin}^2(\frac{a q_i}{2}) \nonumber \\
s &=& \sqrt{\sum_i p_i^2+p_5^2}.
\end{eqnarray} 
With this notation, the eigenvalue problem takes the same form for either discretization; we will we drop the superscript differentiating the two since everything that follows applies equally to both cases. The Hamiltonian in this basis is then
 \begin{align}
H=\left(
\begin{matrix} 
   m_{\text{eff}} \mathbf{1}_2 & \vec{\sigma} \cdot \vec{p}\\
   \vec{\sigma}\cdot \vec{p} & -m_{\text{eff}} \mathbf{1}_2
\end{matrix}
\right),
\label{eqn:HelicityModeEigensystem}
\end{align}
which has eigenvalues $E_\pm=\pm\sqrt{m_{\text{eff}}^2+\vec{p}^2}$, where the positive (negative) eigenvalues corresponds to (anti) particles. The corresponding eigenvectors are given as
\begin{align}
\label{eqn:Wilson_Eigenmodes}
u^h(p)&=\sqrt{\frac{2E_+(E_+-m_{\text{eff}})}{p^2}}\left(
\begin{matrix} 
\phi^{(h)}(p)\\
h\frac{|E|-m_{\text{eff}}}{|p|}\phi^{(h)}(p)
\end{matrix}
\right) \nonumber \\
v^h(p)&=\sqrt{\frac{2E_-(E_- -m_{\text{eff}})}{p^2}}\left(
\begin{matrix} 
\phi^{(h)}(-p)\\
-h\frac{|E|+m_{\text{eff}}}{|p|}\phi^{(h)}(-p).
\end{matrix}
\right),
\end{align}
Since the Hamiltonian, \Eq{eqn:HelicityModeEigensystem}, commutes the helicity operator, the eigenvectors of the Hamiltonian are simultaneously eigenvectors of the helicity operator. 
We then choose $\phi$ to be normalized with respect to helicity $\frac{\vec{p} \cdot \vec{\sigma}}{|\vec{p}|}$, so the index $h$ is the helicity and takes values $h=\pm 1$. Now we solve for the $\phi$. First, 
if $(p_x,p_y)\in \{0,N_x/2\} \times \{0,N_y/2\}$, then
\begin{align}
\phi^+(p)=(1,0)^T\\
\phi^-(p)=(0,1)^T
\end{align}
otherwise
\begin{align}
\phi^h(p)=\frac{1}{\sqrt{1+\frac{(p_z-h|\vec{p}|)^2}{p_x^2+p_y^2}}}\left(\begin{matrix}
1\\
-\frac{p_z-h\sqrt{p_x^2+p_y^2+p_z^2}}{p_x-ip_y}
\end{matrix}\right)
\end{align}
For the case $(p_x,p_y,p_z) \in \{0,N_x/2\} \times \{0,N_y/2\} \times \{0,N_z/2\}$, where the linear momentum term vanishes, for $m_{\text{eff}}>0$
\begin{align}
u^h(p)=\left(
\begin{matrix} 
\phi^h(0)\\ 0
\end{matrix}
\right),~~
v^h(p)=\left(
\begin{matrix}
0 \\
\phi^h(0)
\end{matrix}
\right),
\end{align}
while for $m_{\text{eff}}<0$ we have
\begin{align}
u^h(p)=\left(
\begin{matrix}
0 \\
\phi^h(0)
\end{matrix}
\right),
~~
v^h(p)=\left(
\begin{matrix} 
\phi^h(0)\\
0
\end{matrix}
\right).
\end{align}

While this is most obvious in the last case, the orthogonality conditions
\begin{align}
u^\dagger_{q,\lambda}u_{q,\lambda '}=\delta_{\lambda,\lambda'}\\
v^\dagger_{q,\lambda}v_{q,\lambda '}=\delta_{\lambda,\lambda'}\\
u^\dagger_{q,\lambda}v_{q,\lambda '}=v^\dagger_{q,\lambda}u_{q,\lambda '}=0.
\end{align}
are held for all eigenvectors. We have now constructed the helicity eigenmodes for the free Wilson and overlap Dirac Hamiltonian.  

\section{Convergence study of net axial charge for Wilson and Overlap fermions}
\label{app:convergence}
noticeable
In this appendix we will discuss finite size effects and convergence of our Wilson (see Sec.~\ref{sec:Wison_Fermion_Real_Time}) and overlap (see Sec.~\ref{sec:Overlap_Fermion_Real_Time}) lattice fermions, as well 
as compare the properties of two fermion discretizations. In order to be able to concentrate on the chiral properties of the fermions as a function of volume, improvement, and discretization, we will only consider 
the single sphaleron transition introduced in Sec.~\ref{sec:su2_links}. We keep $\rsph/a=6$ fixed for all simulations and consider only isotropic lattices in this section, and will keep the Wilson r-parameter fixed at $r_w=1$ for all comparisons. In this section we work in the nearly massless limit for the Wilson fermions ($m \rsph=1.9\cdot 10^{-2}$) and the massless limit for overlap fermions, so the integrated anomaly equation reduces to \Eq{eqn:integrated_anomaly}.
We have previously shown for Wilson fermions how both the unintegrated (see Fig.~\ref{fig:profiles_B0}) and integrated (Figure 3 in~\cite{Mueller:2016ven}) anomaly equation are maintained as a function of mass. 
For the Wilson fermions, we first pick a volume, $N^3=16^3$, and study the total axial charge created as a function of time for various levels of operator improvement, as was discussed in Sec.~\ref{sec:Wison_Fermion_Real_Time}. 
This is plotted in Fig.~\ref{sfig:Wilson_Imp_Comp}. We can clearly see that at Leading Order (LO), the standard unimproved Wilson fermion formulation, there is significant deviation, at the 25\% level, from the Chern Simons term 
$-2 \Delta N_{CS}$, which is quantified in the lower panel of Fig.~\ref{sfig:Wilson_Imp_Comp}. However, upon going to one level of improvement, Next to Leading Order improvement (NLO), we see that this 
disagreement disappears. At Next to Next to Leading Order (NNLO) improvement, we see no noticeable difference from NLO, and thus see that our improvement scheme has converged. In practice, we find that 
in all cases in our current study, NLO is sufficient and nothing additional is gained by going to NNLO. 

Now we need to understand how important finite volume effects are in our study. This is shown in Fig.~\ref{sfig:Wilson_Volume_Comp}. Here we look at the axial charge generated by NLO improved Wilson fermions for 
three volumes. It is clear from the lower panel of Fig.~\ref{sfig:Wilson_Volume_Comp} that for $N=12=2 \rsph/a$, there are clear finite volume effects that lead to large oscillations of the $J^0_a$ around the 
sphaleron transition from \Eq{eqn:integrated_anomaly}. This is then subsequently improved by going to a volume $N=16=2.67 \rsph/a$, where we can see noticeable improvement. Similar results for $N=32=5.34 \rsph/a$ indicate convergence to the infinite volume limit.

However, we should note that this is only for 
resolving the creation of axial charge from a single localized sphaleron transition. To look at charge transport as a function of time, like we studied in Sec ~\ref{sec:CME}, we need even larger volumes, especially 
in the magnetic field direction. Typically we choose a spatially anisotropic lattice, where the transverse length is $N_{trans} \geq 2 \rsph/a$, while along the direction of the magnetic field $N_z \gg 2 \rsph/a$ 
(a typical choice is $N^3=16^2 \times 32 - 24^2 \times 64$). Moreover, the transverse size of the lattice has to be large enough to accommodate the cyclotron orbits of charged particles. In practice this constraint limits the available magnetic field strength to larger magnetic flux quanta.

Next, for the overlap fermions, we proceed in the same manner. Instead of improving the Wilson kernel, we vary the domain wall height $M$ for a fixed isotropic lattice $N=16$. As we see in Fig.~\ref{sfig:Overlap_Anomaly_Difference}, values in the range of $M \in [1.4,1.6)$ give the best results; we choose $M=1.5$. We have verified that the volume dependence of the currents for the overlap is similar to the Wilson fermions with NLO improvement, which is evident from \Fig{fig:HMS_COMP}. 

In summary, for Wilson fermions, NLO improvement is necessary and sufficient to accurately reproduce the anomaly. At this level, we find that it gives comparable results to the overlap fermions, which we find 
that for a well tuned domain wall mass $M$ we can reproduce the anomaly relation even on reasonably small lattices. Additionally, we find that for spatial lattice sizes of $N=2~\rsph/a$, finite volume effects 
are somewhat noticeable, but seem to be completely under control for lattice size $N > 2 \rsph/a$. This will serve also as crucial input for how fine to make one's lattice for future studies with more 
realistic gauge field configurations, where the size sphalerons is set by physical scales of the problem.

  \begin{figure}
  \includegraphics[width=.95\linewidth]{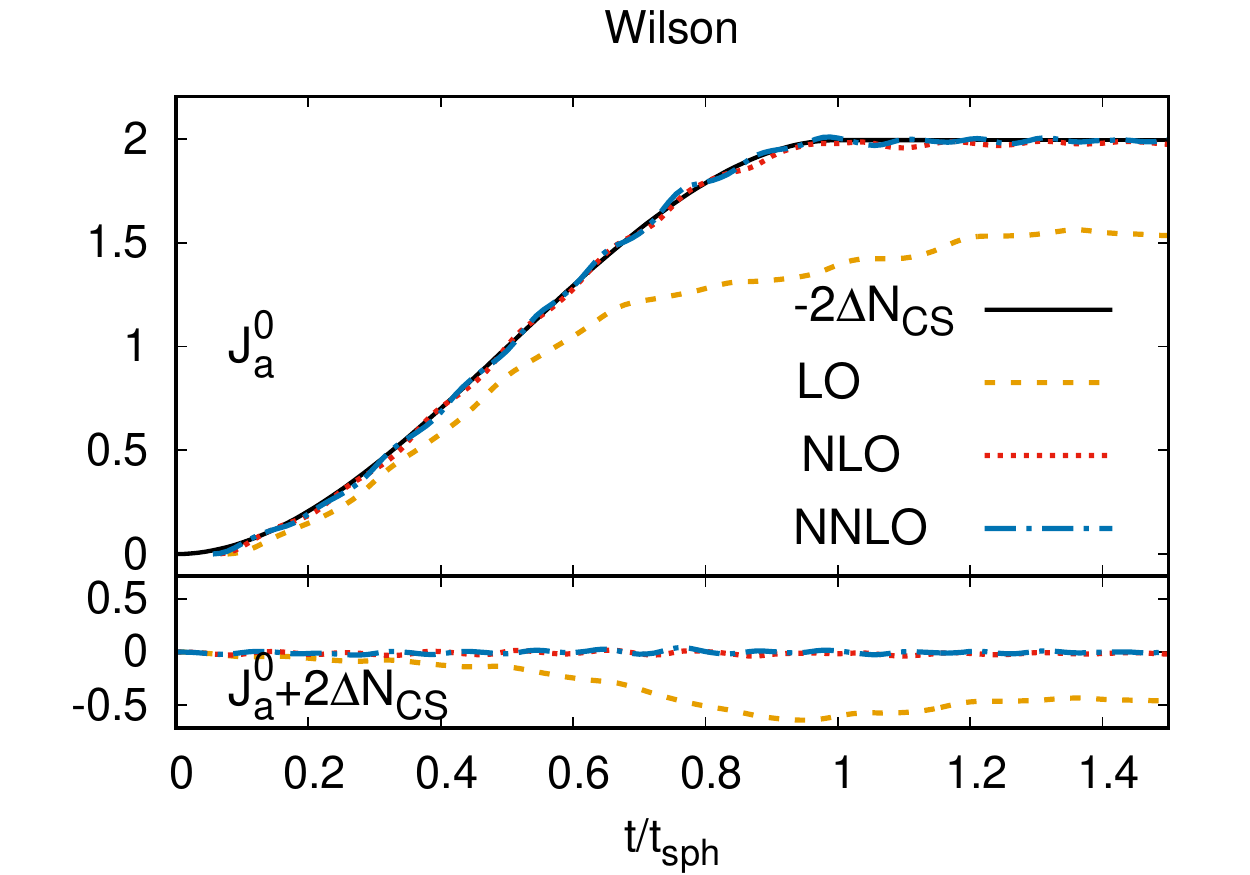}
  \caption{A comparison of the net axial charge generated during a sphaleron transition for a fixed volume of $N=16$ using $m\rsph=1.9 \cdot 10^{-2}$ Wilson fermions with different operator improvements. Top: Already at NLO we see that the net axial charge tracks $\Delta N_{CS}$ due to the sphaleron transition. Bottom: Deviations from \Eq{eqn:integrated_anomaly} are shown.}
  \label{sfig:Wilson_Imp_Comp}
  \end{figure}
  
\begin{figure}
  \includegraphics[width=.95\linewidth]{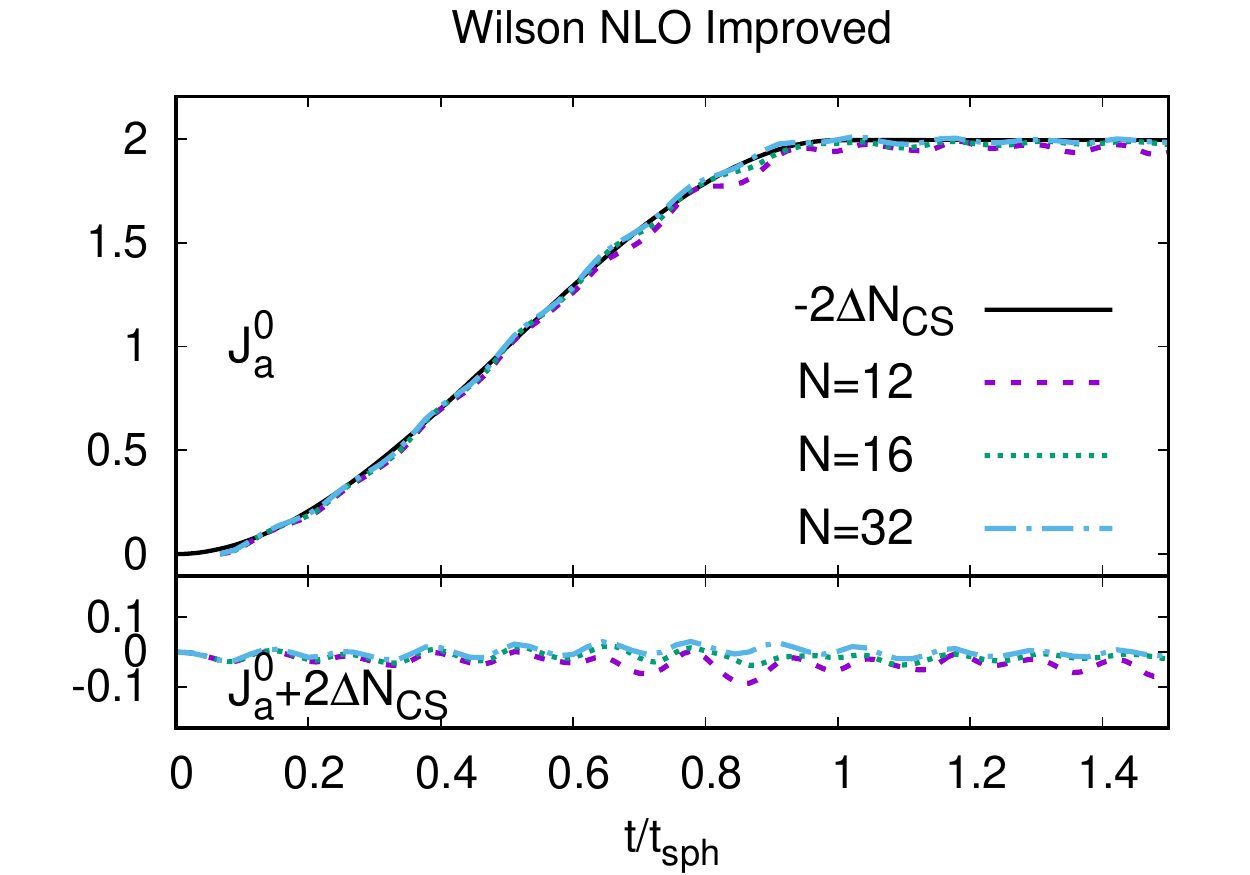}
  \caption{A comparison of the net axial charge generated during a sphaleron transition using $m\rsph=1.9 \cdot 10^{-2}$ improved Wilson (NLO) fermions for different lattice volumes. Top: At N=16 and beyond the net axial charge tracks $\Delta N_{CS}$ due to the sphaleron transition. Bottom: Deviations from \Eq{eqn:integrated_anomaly} are shown.}
  \label{sfig:Wilson_Volume_Comp}
  \end{figure}

    \begin{figure}
  \includegraphics[width=.95\linewidth]{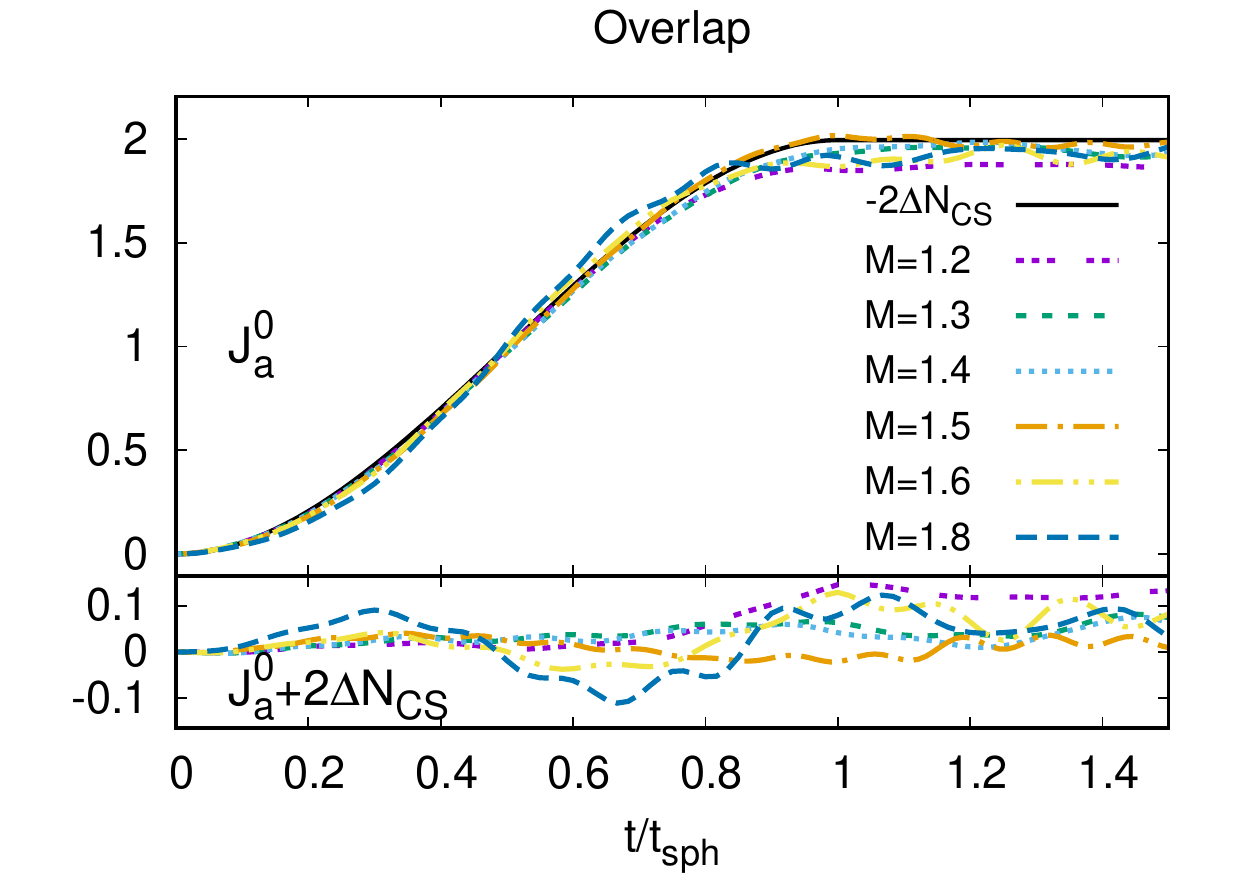}
  \caption{A comparison of the net axial charge generated during a sphaleron transition using massless overlap fermions for different domain wall heights $M$ at a fixed lattice volume $N=16$. Top: For $M\in[1.4,1.6)$, the net axial charge tracks $\Delta N_{CS}$ due to the sphaleron transition. Bottom: Deviations from \Eq{eqn:integrated_anomaly} are shown.}
  \label{sfig:Overlap_Anomaly_Difference}
  \end{figure}

\section{Derivation of the overlap Hamiltonian}
\label{app:Ovhamiltonian}
In this appendix, we outline our construction of the overlap Hamiltonian in 3+1D Minkowski spacetime, applicable for real-time lattice gauge theory simulations. The spatial overlap operator for one massless quark flavor is defined as
\begin{eqnarray}
-i\slashed{D}_{ov}&=& M\Big(\mathbf{1}+\gamma_5 \frac{Q}{\sqrt{Q^2}}\Big),
\end{eqnarray}
where a suitable choice of the kernel $Q$ is 
\begin{eqnarray}
Q\equiv -i\gamma_5 \slashed{D}_W(M),
\end{eqnarray} 
with $-i \slashed{D}_W(M)$ being the massless Wilson Dirac operator in 3+1D Minkowski spacetime. Here the parameter $M\in[0,2)$ can be interpreted 
as the height of the domain wall or the defect that localizes the chiral fermions on 4D Euclidean spacetime starting from a 5D domain wall formalism~\cite{Kaplan:1992bt}. 

In order to derive the real-time evolution of fermion modes $\psi$ with mass $m=0$ and at any instant of time $t$, we solve the overlap Dirac equation on the lattice, $-i\slashed{D}_{ov} \psi=0$, where
\begin{eqnarray}
\label{eqn:OverlapOperatorDw}
-i\slashed{D}_{ov} \psi &=& M\Big[\mathbf{1}+\frac{-i \slashed{D}_W(M)}{\sqrt{{\gamma_5 (-i \slashed{D}_W(M))} \gamma_5 (-i \slashed{D}_W(M))}}\Big]\psi \nonumber \\
\end{eqnarray}
In the temporal gauge and furthermore choosing the lattice spacing along temporal direction to be fine enough than the other relevant scales in the operator, such that $a_t \ll M, a_s$, 
the dimensionless overlap operator is simply
\begin{eqnarray}
\label{eqn:OverlapOperatorDw1}
-i\slashed{D}_{ov}=M\Big[\mathbf{1}+\frac{-i \slashed{D}_W^s-i a_t \slashed{\partial}_t -M}{\sqrt{ \slashed{D}_W^s \slashed{D}_W^{s\dagger} +a_t^2 \partial_t^2 +M^2}}\Big]~. 
\end{eqnarray}
If we perform an expansion in powers of $a_t$ and keep terms which are leading order in $a_t$, we get the RHS of \Eq{eqn:OverlapOperatorDw1} to be,
\begin{eqnarray}
\label{eqn:OverlapOperatorDw2}
M\Big[\mathbf{1}+\frac{-i \slashed{D}_W^s -M}{\sqrt{ \slashed{D}_W^s \slashed{D}_W^{s\dagger} +M^2}}+
\frac{-i a_t \slashed{\partial}_t }{\sqrt{ \slashed{D}_W^s \slashed{D}_W^{s\dagger} +M^2}}\Big] 
\end{eqnarray}
In the denominator of the second term of \Eq{eqn:OverlapOperatorDw2}, the domain-wall height scales as $1/a_s$, whereas the spatial Wilson-Dirac operator scales as linear power in $a_s$, therefore 
the overlap operator in \Eq{eqn:OverlapOperatorDw2} simply reduces to
\begin{eqnarray}
\label{eqn:OverlapOperatorDw3}
-i\slashed{D}_{ov}=-i a_t\slashed{\partial}_t+M\Big[\mathbf{1}+\frac{-i \slashed{D}_W^s -M}{\sqrt{ \slashed{D}_W^s \slashed{D}_W^{s\dagger} +M^2}}\Big]
\end{eqnarray}
The overlap Dirac equation in \Eq{eqn:OverlapOperatorDw} can be then simply written as a time evolution equation of the form,
 \begin{equation}
\label{eq:DiracEquationOv}
i \gamma^{0} \partial_t \psi=-i\slashed{D}^{s}_{ov} \psi
\end{equation} 
where $-i\slashed{D}^{s}_{ov}$ is the spatial overlap operator given by
\begin{eqnarray}
\label{eqn:OverlapOperator3d}
-i\slashed{D}_{ov}^s&=&M\Big[\mathbf{1}+\frac{-i \slashed{D}_W^s(M)}{\sqrt{{\gamma_5 (-i \slashed{D}_W^s(M))} \gamma_5 
(-i \slashed{D}_W^s(M))}}\Big].\nonumber \\
\end{eqnarray}
\Eq{eq:DiracEquationOv} is the analogue of the corresponding evolution equation with Wilson fermion discretization given in \Eq{eq:DiracEquationW}. Using $\gamma_5$ and $\gamma_0$ hermiticity of 
$-i\slashed{D}_{W}^s$, we can recast \Eq{eq:DiracEquationOv} as a Hamiltonian equation with the overlap Hamiltonian in 3D Minkowski space for massless fermions defined as,
\begin{eqnarray}
\label{eqn:OverlapOperatorhapp}
H_{ov}=-i\gamma_{0} \slashed{D}_{ov}^s&=&M\Big(\gamma^{0}+\frac{H_W(M)}{\sqrt{ H_W(M)^2}} \Big),
\end{eqnarray}
where $H_W$ is the Wilson Hamiltonian defined in \Eq{eqn:WilsonHamiltonian} but with $C_n=0$ for $n\geq2$ and the mass $m$ being replaced by the negative of the domain wall height $M$.

\section{Construction of topologically non-trivial lattice map}
\label{app:mapping}

Below we describe the construction of a topologically non-trivial gauge transformation, employed in Sec.~\ref{sec:su2_links} for the construction of our handmade sphaleron transition. We first note that in the continuum, gauge transformations  $G: \mathbb{R}^3\cup\{\infty\} \to SU(2)\simeq S^{3}$ parametrized according to
\begin{align}\label{eq:gaugeobject}
G_\xs=\alpha_0(\xs)\mathbf{1}+i\alpha^a(\xs) \sigma^a,\qquad \alpha_0^2+\mathbf{\alpha}^2=1\;.
\end{align}
where  $a=1,\cdots,3$ and $\alpha^2=\alpha_{a}\alpha^{a}$ can be classified according to the homotopy classes $\pi_{3}(\mathbb{R}^3\cup\{\infty\})\simeq \mathbb{Z}$ characterized by the topological winding number or Brouwer degree $deg(G)$ of the map. Even though on a lattice with periodic boundary conditions the corresponding map is from the three torus $\mathbb{T}^3$ to the gauge group, the homotopy classes $\pi_{3}(\mathbb{R}^3\cup\{\infty\})\simeq Z$ and $\pi_{3}(\mathbb{T}^3)\simeq \mathbb{Z}$ are identical and the same classification scheme applies. Our strategy to construct a topologically non-trivial lattice map is to perform a multi-step mapping; first from the set of lattice points to the three torus, then to one-point compactified real space, and finally to the gauge group, such that
\begin{align}
\mathbf{x}  \to \mathbf{x}_{T} \to \mathbf{x}_\mathbb{R} \to  \mathbf{\alpha}
\end{align}
with $\mathbf{x} \in \{0,\cdots,N_x-1\}\times \{0,\cdots,N_y-1\} \times \{0,\cdots,N_z-1\}$, $\mathbf{x}_{T} \in T^{3}$, $\mathbf{x}_\mathbb{R} \in \mathbb{R}^3\cup\{\infty\}$ and $\mathbf{\alpha} \in S^{3}\simeq SU(2)$.  Since we wish to obtain a non-trivial result $(G_{\xs}\neq1)$ only on a characteristic scale $\rsph$ around the center $(x,y,z)=(N_x/2,N_y/2,N_z/2)$, we perform a distorted map of the lattice points to real space, given explicitly by the following steps
\begin{align}
x^i_T &= 2\pi\frac{\text{arctan}\left( \frac{x^i-N^i/2}{\rsph} \right) - \text{arctan}\left( \frac{-N^i/2}{\rsph} \right)}{\text{arctan}\left( \frac{N^i/2}{\rsph} \right) - \text{arctan}\left( \frac{-N^i/2}{\rsph} \right)},
\end{align}
where the denominator explicitly ensures a smooth profile at the periodic edges, and subsequently
\begin{align}
x^i_\mathbb{R}&=\tan\left[ \frac{1}{2}\left(x^i_T-\pi\right)\right],
\end{align}
which identifies a unique point in $\mathbb{R}^3\cup\{\infty\}$ with each lattice site. Since, in order to ensure that the final map between $\mathbb{R}^3\cup\{\infty\}$ and $\mathbf{\alpha} \in S^{3}\simeq SU(2)$ has a non-vanishing degree, it is sufficient to require the map to be surjective; we simply choose the final mapping to be given by a stereographic projection, such that
\begin{align}
\alpha_0&=\frac{\mathbf{x}_\mathbb{R}^2-1}{\mathbf{x}_\mathbb{R}^2+1}\nonumber\\
\alpha^a&=(\alpha_0-1)\cdot {x}_\mathbb{R}^i\delta^{ia}.
\label{eq:mapstereo}
\end{align}
Based on the explicit formula for the Brouwer degree, it is straightforward to verify that this map has topological winding number equal to unity.

\end{document}